\documentclass[11pt]{article} 
\usepackage{graphicx}						
\usepackage{amssymb, amsmath, amsthm}
\usepackage{amsfonts}
\newcommand{\argmin}{\mathop{\mathrm{arg\,min}}}
\usepackage{bm}
\usepackage{bbm}
\usepackage[ruled,linesnumbered, vlined, noend]{algorithm2e}
\usepackage{xcolor}

\title{Improving Fairness in Criminal Justice Algorithmic Risk Assessments Using Optimal Transport and Conformal Prediction Sets\thanks{Cary Coglianese and Sandra Mayson provided many insightful suggestions for legal conceptions of fairness and the prospect for criminal justice reform. We also received very useful feedback from a group of researchers at MIT and Harvard who work on causal inference. In that regard, a special thanks goes to Devavrat Shah. Thanks also go to three thoughtful reviewers.}}
\author{Richard A. Berk \\ \small{University of Pennsylvania}\\
Arun Kumar Kuchibhotla \\  \small{Carnegie Mellon University} \\ Eric Tchetgen Tchetgen \\ \small{University of Pennsylvania}}
\date{}
\begin{document}
\maketitle
\begin{abstract}
In the United States and elsewhere, risk assessment algorithms are being used to help inform criminal justice decision-makers. A common intent is to forecast an offender's ``future dangerousness.'' Such algorithms have been correctly criticized for potential unfairness, and there is an active cottage industry trying to make repairs. In this paper, we use counterfactual reasoning to consider the prospects for improved fairness when members of a disadvantaged class are treated by a risk algorithm as if they are members of an advantaged class. We combine a machine learning classifier trained in a novel manner with an optimal transport adjustment for the relevant joint probability distributions, which together provide a constructive response to claims of bias-in-bias-out. A key distinction is made between fairness claims that are empirically testable and fairness claims that are not. We then use confusion tables and conformal prediction sets to evaluate achieved fairness for estimated risk. Our data are a random sample of 300,000 offenders at their arraignments for a large metropolitan area in the United States during which decisions to release or detain are made. We show that substantial improvement in fairness can be achieved consistent with a Pareto improvement for legally protected classes.

\pagebreak

\noindent \textbf{Keywords}

Risk Assessment; Fairness; Risk Algorithms; Machine Learning; Optimal Transport; Conformal Prediction Sets
\end{abstract}	

\section{Introduction}
The goal of fair risk algorithms in criminal justice settings remains a high priority among algorithm developers and the users of those algorithms. The literature is large, scattered, and growing rapidly, but there seem to be three related conceptual clusters: definitions of fairness and the trade-offs that necessarily follow (Kleinberg et al., 2017; Kroll et al., 2017, Corbett-Davies and Goel, 2018), claims of ubiquitous unfairness (Harcourt, 2007; Star, 2014; Tonrey, 2014; Mullainathan, 2018), and a host of proposals for technical solutions (Kamiran and Calders, 2012;  Hardt et al., 2016; Feldman et al. 2015; Zafer et al., 2017; Kearns et al., 2018; Madras et al., 2019; Lee et al., 2019; Johndrow and Lum, 2019; Romano et al., 2019; Skeem and Lowenkamp, 2020). There are also useful overviews that cut across these domains (Berk et. al. 2018; Baer et al., 2020; Mitchell et al., 2021)

In this paper, we examine risk assessments used in criminal justice settings and propose a novel adjustment to further algorithmic fairness. Because the outcome of interest is categorical, we concentrate on algorithmic classifiers. Unlike most other work, the methods we offer are in part a response to a political climate in which appearances can be more important than facts, and political gridlock is a common consequence. To help break the gridlock, we seek a remedy for algorithmic unfairness that is politically acceptable to stakeholders. In so doing, we take a hard look at what risk algorithms realistically can be expected to accomplish. 

A recent paper by  Berk and Elzarka (2020) provides a good start by proposing an unconventional way a fair algorithm could be trained. However, their approach lacks the formal structure we offer, which, in turn, solves problems that the earlier work cannot. Building on a foundation of machine learning, optimal transport (H\"utter and Rigollet, 2020; Ni et al., 2021), and conformal prediction sets (Vovk et al., 2005; 2009), we suggest a justification for risk algorithms that treats members of a less ``privileged,'' legally protected class as if they were members of a more ``privileged,'' legally protected class. We use Black offenders and White offenders at their arraignments to illustrate our approach with the forecasting target an arrest for a crime of violence. Many will claim that Black offenders are members of a less privileged, legally protected class and White offenders are members of a more privileged, legally protected class. Less freighted terms are ``disadvantaged'' and ``advantaged'' respectively. More will be said about the definition of ``protected class'' and the legal recourse available in the next section. 

As a first approximation, if the performance of a risk algorithm can serve as an acceptable standard for the class of White offenders, it arguably can serve as an acceptable standard for the class of Black offenders. This helps to underscore that we propose altering only how \emph{different protected classes are treated by a risk classifier}. We are not suggesting that a single risk algorithm can make fundamental change throughout the criminal justice system; the reach of a criminal justice risk algorithm is far more modest. We also are not suggesting that the risk classifier must be state-of-the-art or even without significant defects. It must just be the customary procedure for Whites in a particular jurisdiction. 

Like many who have written on algorithmic fairness, we consider fairness for legally protected classes; we focus on \emph{group} fairness. Consistent with most work on fair algorithms, we take the definition of a protected class as given and contingent upon the application context (Coston et al., 2019: Section 2.2). As a statutory and legal matter, how groups of people come to be defined as a protected class is a complicated historical and political process often weakly tethered to first principles (Allen and Crook, 2017). The issues are well beyond the scope of this paper.

Whatever the processes responsible for creating legally protected classes, there can be legitimate fairness concerns about treating Black offenders as if they are White. To some, we are proposing inequality of treatment. But here too, the issues are complex. There is no universal proscription of using race to formulate fairness remedies. On the use of race to counter discriminatory practices, Kim notes (2020:5) that ``As numerous scholars have pointed out, the law does not categorically prohibit race-consciousness.'' Context matters. Moreover, we argue later that the protected class of Black offenders is made better off while the protected class of White offenders is not made worse off. We can achieve a form of Pareto improvement at the level of legally protected classes. This is quite different from the manner in which controversial interventions such as affirmative action for college admissions are often characterized. Some protected classes are said to made better off by the admissions reforms while others are said to be made worse off (Regents of the University of California v. Bakke, 1978).

We also respond constructively to a long standing ethical quandary in U.S. criminal justice (Fisher and Kadane, 1983) commonly neglected in recent overviews of fairness (Baer et al., 2020; Mitchell et al., 2021). Should adjustments towards racial fairness use the treatment of White offenders as the baseline, the treatment of Black offenders as the baseline, or some compromise between the two? Although in principle, equality may be achieved using any shared baseline, the protected classes made better off and the protected classes made worse can differ dramatically. Unless an acceptable baseline for all is determined, there likely will be no agreement on how fairness is to be achieved. Moreover, when the baseline is not addressed along with adjustments toward fairness, one can arrive at fair results in which all protected classes are made equally worse off. It is difficult to imagine that stakeholders would find such a result palatable.

Whatever the baseline, a common and curious supposition is that any disparity in treatment or outcome between different protected groups is unfair and even discriminatory (Mason, 2019). Gender provides an especially clear example. Men are disproportionately over-represented in prisons compared to women. But throughout recorded history, men disproportionately have committed the vast majority of violent crimes. Is the gender disparity in imprisonment explained solely by unfairness? For criminal justice more broadly, finding comprehensive explanations for racial disparities likewise is challenging (Hudson, 1989; Yates, 1997). We do not pretend to have a definitive resolution but offer instead what we hope is a politically acceptable approach. 

Finally, treating Black offenders as if they are White, complicates how the appropriate risk estimands should be defined. Necessarily, counterfactuals in some form are being introduced because Black offenders cannot be White offenders, and a fair risk algorithm does not make fair all criminal justice decisions and actions that follow. We discuss these issues in the context of estimates produced by a classifier and also with confusion tables and conformal prediction sets. In so doing, we introduce counterfactual estimands to help clarify distinctions between fair risk assessment procedures and fairness in the criminal justice system more generally. The two are often conflated. A risk procedure ends with the output of a risk tool. Everything that follows, whether decisions or actions, are features of the criminal justice system beyond the risk algorithm. 

In section 2, we discuss definitions of fairness in the statistics and computer science literature commonly associated with criminal justice risk assessment and introduce two key concepts: internal fairness and external fairness. Section 3 summarizes the methods we use to improve fairness in criminal justice risk assessments: how a classifier should be trained, how to make comparable joint distributions of the data from different protected classes, and how to gauge fairness using conformal prediction sets. These methods are discussed more formally in Appendices A through E. Section 4 describes the data to be analyzed: 300,000 individuals who are arraigned shortly after an arrest and subsequently released. Section 5 discusses the results; can post-release arrests for a crime of violence be usefully and fairly forecasted? In section 6, we return in depth to the empirical results shown as conformal prediction sets. Section 7 suggests some possible generalizations, and section 8 provides conclusions.

\section{Conceptions of Algorithmic Fairness for Protected Classes}

We are concerned in this paper with group fairness. Legally protected classes are the groups. Just as in formulations of group fairness, protected classes have a different conceptual status from their constituent individuals.

\subsection{Some Background on Legally Protected Classes}

Under the U.S. Civil Rights Act of 1964, and subsequent federal anti-discrimination statutes, a ``protected class'' is a collection of people that can have explicit protection against discrimination, consistent with the 5th and 14th Amendments to the United States Constitution. Protected classes in anti-discrimination law are called ``suspect classifications'' in constitutional law. Examples of attributes that characterize protected classes are race, religion, national origin, and gender.  

The protection takes the form of judicial scrutiny, which can vary in intrusiveness, but only opens the door to an examination of the setting, the specific facts, and how well tailored the use of class membership may be (Kim, 2020). There are no foregone conclusions. From a litigation perspective, moreover, class membership is not automatic. For example, a particular black employee is not a member of a protected class until the employee formally alleges, say, disparate treatment in wages by his or her employer. The class membership persists only until the end of legal dispute.\footnote
{
The distinction between a protected class and its potential members has enormous practical implications as well. For example, in the death penalty case of McCleskey v. Kemp (481 U.S. 279 (1987)), the U.S. Supreme Court ruled that despite evidence of a racially disproportionate impact of the Georgia death penalty, there was no showing that Warren McCleskey, who was Black, experienced conscious, deliberate bias as an individual. The death penalty sentence was upheld. In this paper, we consider how one might determine whether there is a racially disproportionate impact of a risk algorithm and if so, how effective repairs can be made. We do not consider how one might determine whether an individual has been subject to algorithmic unfairness because of race, and if so, what reparations should be provided. These are conceptually and empirically a very different issues.  
}

This leads to some semantic ambiguities. For example, race is commonly called a protected class that can include several different races such as Whites, Blacks and Asians. It is then also common to speak of particular Whites, or Blacks, or Asians, as a protected class when they seek legal redress for discrimination. In context, however, there should be no confusion. 

\subsection{Fairness Definitions}

Even for legally protected classes, there is no constitutional guidance for definitions of algorithmic fairness, in part because jurisprudence is still trying to catch up. There are no prescribed empirical methods. And, there is not even a common language to address the issues (Mason, 2019; Huq, 2019). Nevertheless, researchers have proposed a variety of algorithmic fairness types. 

The definitions to follow arise directly from confusion tables and are easily translated into many common fairness typologies employed in risk assessment (Kleinberg et al., 2017; Kroll et al., 2017; Berk et al., 2018, Baer et al., 2020; Mitchell et al., 2021). Other definitions briefly are considered in due time. Anticipating our later data analyses, we use White criminal justice offenders and Black criminal justice offenders at their arraignments as illustrations of the protected classes for which fairness is sought. Fairness centers on their algorithmic forecasts of risk. Is the \emph{algorithmic output itself} fair?
\begin{itemize}
\item
\textit{Prediction parity} -- The predictive distributions across protected classes are the same. Prediction parity can be estimated with test data by, for example, the proportions of Black offenders or White offenders forecasted to be arrested after an arraignment release. Prediction parity is sometimes called demographic parity. 

Prediction parity is judged by the risk tool output alone, not the decisions that follow, any subsequent actions or occurrences, or the true but unknown outcomes. An important implication is that the actual outcome class to be forecasted (e.g., an arrest) has no role in the definition of prediction parity.  As a result, prediction parity has been criticized as unsatisfactory and even irrelevant (Hardt et al., 2016). Yet, an absence of prediction parity may be linked to ``mass incarceration," which in practice cannot easily be disregarded. Moreover, requiring the inclusion of the actual outcome class in fairness definitions leads to challenges we discuss shortly.
\item
\textit{Classification parity} -- The false positive rates and false negative rates are the same across protected classes. A false positive denotes that a risk algorithm incorrectly classified a case with a negative class label as a case with a positive class label. A false negative denotes that a risk algorithm incorrectly classified a case with a positive class label as a case with a negative class label.\footnote
{
Which outcome class is a positive and which is a negative is determined by the subject matter or policy being addressed. In the analysis to follow, an arrest for a violent crime is a negative, and the absence of such an arrest is a positive.
}
The false positive rate and a false negative rate are the respective probabilities that the algorithm classifies outcomes erroneously.  When there are more than two outcome classes, classification parity follows from the same reasoning, but there are no common naming conventions.

Ideally, false positive and false negative rates are estimated with test data. Classification error for a particular outcome class, such as an arrest, is the proportion of subjects erroneously classified as not arrested among all who actually were arrested. If an arrest is the positive class, for example, one has an estimate of the false negative rate. More formally, 

\begin{equation}\label{eq:classification-error-fair}
\begin{split}
&\mbox{Estimated Classification Error (for an arrest)}\\
&\quad:= \frac{\sum_{i\in \mathrm{test}} \mathbbm{1}\{\widehat{Y}_i \neq Y_i, Y_i = \mbox{arrest}\}}{\sum_{i\in \mathrm{test}} \mathbbm{1}\{Y_i = \mbox{arrest}\}}.
\end{split}
\end{equation}
$Y_i$ is the true outcome for subject $i$ in test data, and $\widehat{Y}_i$ is the forecasted, test data outcome from the trained classifier (i.e., usually the outcome with the highest estimated probability). The classification error in~\eqref{eq:classification-error-fair} is an estimator of $\mathbb{P}(\widehat{Y} \neq Y|Y = \mbox{arrest})$. Note that one conditions on the true outcome class.

Classification error, whether through false positives or false negatives, has played a central role in fairness discussions by statisticians and computer scientists (Baer et al., 2020). However, it is often irrelevant to stakeholders, who in practice care far more about forecasting accuracy. In real forecasting settings, the actual outcome is unknown, and any subsequent decisions can be primarily informed by the forecasted outcome. Further, emphasizing classification may encourage interpretations akin to the prosecutor's fallacy (Thompson and Schumann, 1987); classification accuracy is used inappropriately to evaluate forecasting accuracy.     
\item
\textit{Forecasting accuracy parity} -- Each outcome class is forecasted with equal accuracy for each protected class. A forecast is incorrect if the forecasted outcome does not correspond to the actual outcome. In contrast to classification parity, one conditions on the forecasted outcome not the actual outcome.

Here too, estimation should be undertaken with test data. The forecasting error for a particular outcome, such as an arrest, is the proportion of subjects erroneously forecasted to not be arrested among all subjects for whom an arrest was the forecast. Formally,
\begin{equation}\label{eq:forecasting-error}
\begin{split}
&\mbox{Estimated Forecasting Error (for an arrest)}\\ 
&\quad:= \frac{\sum_{i\in \mathrm{test}} \mathbbm{1}\{\widehat{Y}_i \neq Y_i, \widehat{Y}_i = \mbox{arrest}\}}{\sum_{i\in \mathrm{test}} \mathbbm{1}\{\widehat{Y}_i = \mbox{arrest}\}}. 
\end{split}
\end{equation}
The notation $Y_i$ and $\widehat{Y}_i$ is the same as for classification parity. Implemented with test data, forecasting error~\eqref{eq:forecasting-error} is an estimator of $\mathbb{P}(\widehat{Y} \neq Y|\widehat{Y} = \mbox{arrest})$. In practice, only the \emph{forecasted} outcomes are available to inform pending criminal justice decisions about particular individuals. This is different from algorithm training and evaluation using historical data, which typically has actual outcomes. 

In some formulations, achieving forecasting accuracy parity requires that the forecasts are calibrated. For example, suppose a risk tool projects for certain offenders an arrest probability of .68. For calibration to be achieved, the actual arrest probability for all such offenders must also be .68. (Baer at al., 2020). The same reasoning applies to the estimated arrest probabilities for each outcome category. By itself, this criterion is silent on fairness, but it restricts discussion of forecasting accuracy parity to applications in which a risk tool is by this yardstick performing properly. In practice, calibration can be very difficult to attain for criminal justice risk algorithms -- with good reason (Gupta et al., 2020; Lee et al., 2022). This is a major theme to which we will return later.
\item
\textit{Cost Ratio parity} -- The relative costs of false negatives to false positives (or the reciprocal), as defined above, are the same for each protected class. The cost ratio determines the way in which a risk assessment classifier trades false positives against false negatives. Commonly, some risk assessment errors are more costly than others, but the relative costs of those errors should be same for every protected class. The same reasoning applies when there are more than two outcome classes.\footnote
{
These costs are rarely monetized. How would one measure in dollars the ``pain and suffering" of a homicide victim's family or the psychological trauma of neighborhood children who witness a homicide? What matters for the risk algorithm is \emph{relative} costs. For example, failing to accurately identify a prison inmate who after release commits a murder will be seen by many stakeholders as far more costly than failing to accurately identify a prison inmate who after release becomes a model citizen. The cost ratio might 5/1. In practice, the desired relative costs are a policy choice made by stakeholders that, in turn, is built into the risk algorithm. If no such policy choice is made, the algorithm necessarily makes one that can be very different from stakeholder preferences and even common sense. Cost ratios can affect forecasted risk, often dramatically.
} 
\end{itemize}

For reasons that become more apparent shortly, we focus primarily on prediction parity when we consider how one can construct a useful risk algorithm that forecasts arrests for crimes of violence. We aim to have the same proportions of Black and White offenders assigned to each forecasted outcome; there should be prediction parity for these two protected classes. This is a canonical risk algorithm aspiration. To succeed, however, one must overcome several formidable obstacles, some of which are not always appreciated.   

\subsection{Obstacles in Practice}

Practice usually demands that when each kind of parity is evaluated, some form of legitimate comparability is enforced. For instance, if a risk algorithm forecasts that Black offenders are far more likely to be rearrested than White offenders, there is a prediction disparity. But in the training and test data, perhaps Black offenders tend to be younger, and the algorithm may properly recognize that younger individuals are more inclined to commit violent crimes. In this instance, the \emph{class} of Black offenders and the \emph{class} of Whites offenders are not similarly situated. Until this is effectively addressed, one cannot determine whether the disparity between the two protected classes is unfair.

For much of the current fairness literature, whether observational units that comprise protected classes are similarly situated is primarily a technical problem that boils down to methods that adjust for racial confounders, sometimes in a causal model. Typically overlooked is that candidate confounders possess normative as well as causal content, and both affect how confounders are selected. On the closely related notion of culpability, Horder observes (1993: 215) ``... our criminal law shows itself to be the product of the shared history of cultural-moral evolution, assumptions, and conflicts that is the mark of a community of principle.'' As a result, controversies over fairness often begin with stark normative disagreements about what it means to be similarly situated. For example, should an offender's juvenile record matter in determining whether cases are similarly situated? The answer depends in part on how in jurisprudence psychosocial maturity is related to culpability (Loeffler and Chalfin, 2017). 

Normative considerations also can create striking incongruities. Official sentencing guidelines, for example, often prescribe that defendants convicted of the same crimes and with the same criminal records, should receive the same sentences. Under these specific guidelines, such defendants are similarly situated (Ostrom et al., 2003: chapter 1). ``Extralegal" factors such as gender, race, and income are not properly included in that determination. But if ``criminal records'' are significantly a product of gender, race, and income, should they not be extralegal as well?  The extensive literature on fairness cited earlier makes clear that there is no satisfactory answer in sight.
In the pages ahead, we offer a pragmatic way forward. 

There is more clarity about provable trade-offs between certain forms of parity and between parity, accuracy and transparency that are folded into discussions of fairness (Kleinberg et al., 2017; Barocas et al., 2018; Coglianese and Lehr, 2019; Kearns and Roth, 2020; Diana et al., 2021; Mishler and Kennedy, 2021). Hard choices necessarily follow that in applications cannot be made solely for mathematical convenience. Value-driven compromises of various kinds are in practice inevitable. There does not seem to be at this point any technical resolution allowing stakeholders to have it all.

Finally, there is no empirical standard for how small disparities in parity must be for the parity to be acceptable, although most stakeholders agree that small disparities may suffice.\footnote
{
A tactic that has been used to preclude algorithmic risk assessment altogether is to insist on exact parity.
} 
Interpretations of ``small'' will be contentious because harm depends on facts and judgements that are easily disputed. Moreover, there usually is no clear threshold at which some amount of harm becomes too much harm. Similar issues arise across the wide variety of litigation domains (Gastwirth, 2000). The fairness literature has been silent on the matter, and we do not address it here. It is peripheral to our discussion of fairness, but for fair algorithms to be used effectively in practice, a binding resolution is required.

\subsection{Counterfactuals: Internal and External Fairness} 

Despite the challenges, one has estimators for the four kinds of parity that can be employed with the usual test data. Organizing the test data separately into a confusion table for each protected class, one easily can consider the degree to which each kind of parity is achieved. When each kind of parity is examined using a combination of test data and algorithmic output, one is assessing what we call \emph{internal fairness}. 

Prediction parity is a very important special case. Because outcome labels are unnecessary for its definition, one legitimately can examine this form of algorithmic fairness from test data and algorithmic output alone. If the distribution of outcome forecasts is not sufficiently alike across protected classes, one properly may claim that prediction parity has been violated.\footnote
{
There are many simple ways to achieve prediction parity without even using a conventional classifier. For example, at arraignment the presiding magistrate might flip a fair coin. Heads means the offender is high risk. Tails means that the offender is low risk. All offenders would have have the same estimated probability of re-arrest. But the magistrate properly would be accused of procedural capriciousness (Holewinski, 2002). If a magistrate assumed that all offenders were high risk, prediction parity also would be achieved, but this procedure probably would fail by the criterion of arbitrariness. In real settings, methods that create prediction parity must pass jurisprudential muster, comport well with ``self-evident'' truths, and successfully navigate the stakeholder gauntlet. Anything less, can become an academic exercise. 
}

Such claims may really matter. Recall that an absence of prediction parity can be a driving force for mass incarceration. Mass incarceration usually refers to the over-representation of Blacks in the jails and prisons in the United States and is seen by some as modern extension of slavery (Waquant, 2002). It has been a ``hot button'' issue for over a decade (Lynch, 2011), perhaps second only to police shootings in visibility and rancor. 
 
For classification parity, forecasting accuracy parity, and cost ratio parity, one must have the labels for the actual outcomes because those labels are built into their fairness definitions. Typically, each observation in the training and test data has such a label. Yet, we have defined internal fairness such that it depends on test data outcome labels that represent a status quo, which can include racial disparities carried forward as an algorithm is trained and fairness is assessed. For our approach to fairness that treats Black offenders as if they are White, such labels may be especially misleading. We prove in Appendix D that, except for prediction parity, isolating the fairness of a risk algorithm requires untestable causal assumptions that cannot be enforced in practice. These include rank preservation and strong unconfoundedness in how the criminal justice system might treat a person if he or she were of a certain race.\footnote
{
In passing, related issues arise when DAGs and causal reasoning are used to isolate the impact of race on any criminal justice outcome (Baer et al., 2020). 
}

In short, no matter the number of observations or how the data are collected, test data outcome labels for Black offenders, such as an arrest, cannot be assumed to accurately capture the counterfactual of policing in which Blacks are treated the same as Whites. No such world is observed and represented in the data. Yet, accurate information about counterfactuals is a prerequisite for what we call \emph{external fairness}. External fairness is a function for protected classes of the risk algorithm and unobserved fair counterfactuals.

Accurate information about fair counterfactual outcomes is also needed to calibrate risk algorithms properly. In particular, if a risk algorithm achieves a form of fairness by treating Black offenders as if they are White, calibration with respect to the true outcome in test data can be undermined by post-release factors that affect Blacks differently from Whites. For example, a statistical adjustment toward fairness for offenders' prior records may weaken a substantial predictive association, caused by ``overpolicing,'' between longer prior records and an increased chance of a new arrest.  

There are other fairness definitions in the literature for which the concerns effectively are the same. ``Separation'' requires that the forecasted outcome be independent of class membership, given the true outcome (Hardt et al., 2016). ``Sufficiency'' requires that the true outcome be independent of class membership, given the forecasted outcome (Baer et al., 2020). ``Predictive parity'' (not prediction parity) requires that the probability of the true outcome conditional on the forecasted outcome be the same when one also conditions on class membership (Chouldechova, 2017). As before, the fundamental challenge is that the counterfactual outcome for Black offenders is not available in the test data.

Consider, for example, \emph{counterfactual} classification parity as a form of external fairness, where: 

\begin{equation}\label{eq:classification-error}
\begin{split}
&\mbox{Counterfactual Classification Error (for no arrest)}\\ 
&\quad:= \frac{\sum_{i\in \mathrm{test}} \mathbbm{1}\{\widehat{Y}_i \neq Y_i^*, Y_i^* = \mbox{no arrest}\}}{\sum_{i\in \mathrm{test}} \mathbbm{1}\{Y_i^* = \mbox{no arrest}\}},
\end{split}
\end{equation}
and $Y_i^*$ in our formulation denotes the underlying counterfactual outcome for a Black offender treated upon release as a similarly situated White offender. $\widehat{Y}_i$ is the forecast from the trained algorithm. A Black offender's $Y_i^*$ cannot be observed.

Counterfactual outcomes of this sort underscore that all criminal justice risk algorithms necessarily have a circumscribed reach. They are a computational procedures that transform the input with which it is provided into information intended to help inform decisions. One can legitimately ask, therefore, if the algorithmic output by \emph{itself} is fair; is the algorithm ``intrinsically'' fair? A criminal justice risk algorithm is not responsible for the deployment of police assets, the tactics that police employ, easy access to firearms, gang rivalries, and myriad other factors that can affect the likelihood of a post-release arrest. Outcome labels in test data incorporate these factors over which a risk algorithm has no control, and these can produce misleading assessments of classification parity, forecasting accuracy parity and cost ratio parity. One can have an algorithm that itself is fair despite what an analysis using test data shows. Put more strongly, stakeholders are being unrealistic to demand a fair risk algorithm fix widespread inequities in the criminal justice system and the social world more generally.   

It follows that proper evaluations of classification parity, forecasting accuracy, and cost ratio parity may be at this point largely aspirational. Within our approach to fairness and later empirical application, there is no apparent path to sound estimates of the counterfactual world in which race has no role in arrests after an arraignment release such that Black offenders are treated the same as similarly situated, White offenders. One might choose instead to assume that race is unrelated to the many causes of an arrest, but that would be contrary to the overwhelming weight of evidence (Robert Wood Johnson Foundation, 2017; Rucker and Richeson, 2021; Muller, 2021).\footnote
{
Why race is so strongly implicated does not require racial animus by police and other criminal justice agents. To take a tragic example, the numbers of homicides and shootings recently have increased substantially in many American cities. The vast majority of victims are Black. The vast majority of perpetrators are Black. These facts are not a product of racist ``overpolicing,'' racially targeted ``stop-and-frisk'' or racially slanted data. Insofar as the causes are understood, they go to easy access to firearms and long-term structural issues ubiquitous in disadvantaged neighborhoods, perhaps exacerbated by the COVID pandemic (Sorenson et al., 2021). 
}

Other forms of counterfactual reasoning have been proposed for consideration of fairness. For example, Misher and Kennedy (2021; section 2.2) note the potential importance of a race counterfactual somewhat like ours. (i.e., What would happen if an individual's race were different?) Nabi et al. (2019) offer a Directed Acyclic graph (DAG) formulation for fair policies steeped in counterfactuals but requiring assumptions that would be difficult to defend in criminal justice settings. Kusner et al. (2017) provide a DAG framework for examining racial counterfactuals, but it too requires rather daunting assumptions of the sort relied upon in many social science applications (Freedman, 2012). Imai and Jiang (2021) use counterfactual reasoning to define the concept of ``principal fairness,'' which if achieved, subsumes many of the most common kinds of fairness, but requires conditioning on all relevant confounders. The formulation proposed by De Lara et al. (2021), using counterfactual thinking, optimal transport and related tools, is strongly connected to some aspects of our formulation, but makes no distinction between disparities and unfairness such that, for example, there can be no ``bona fide occupational qualifications'' under Title VII of the U.S. Civil Rights Act of 1964, Section 625.\footnote
{
For some occupations, a person's sex, religion, or national origin may be necessary to successfully undertake a job that is a normal activity in a business or enterprise. There can also be legitimate performance requirements as long as all job applicants have an opportunity demonstrate whether they can fulfill those performance requirements. The issues can be subtle. For example, a performance test may misrepresent the true job requirements.
}
 
These interesting papers (see also Mitchell et al., 2020) apply a rich variety of counterfactual ideas to fairness. However, they do not consider many of the foundational fairness concerns raised earlier, such as the need for a fairness baseline or the meaning of ``similarly situated.'' As a result, they currently are some distance from applications in real settings where risk algorithms can directly affect people's lives. They also differ substantially in method and focus from our approach, to which we turn shortly.

Finally, there is a small but powerful literature on fairness for individuals. An early insight is that a fair algorithm at the level of protected classes can be unfair at the level of protected class members. Similarly situated \emph{individuals} are not necessarily treated similarly (Dwork et al., 2012). A solution requires a scale of individual fairness having acceptable mathematical and normative properties. There is apparently no such scale for which there is a meaningful consensus, although some progress has been made (Mukherjee et al. 2020). Perhaps more fundamentally, it may be that class fairness and individual fairness are inherently incompatible (Binns, 2020). Under these circumstances, individual fairness, although an important concern, is not considered further. There currently are far too many unresolved issues that are beyond the scope of this paper. 

\section{Achieving A Fair Criminal Justice Risk Assessment Procedure}

When a criminal justice risk algorithm is trained, the data consist of two parts. There are predictors, and there are outcome labels. The former are used to fit the latter. Both can be responsible if racial disparities are carried forward when a risk algorithm is trained. 

Even when race is not included among the predictors, it is widely understood that all predictors can have racial content (Berk, 2009). For example, the number of prior arrests may be on the average greater for Blacks offenders than for White offenders. Black offenders may also tend to be younger and have been first arrested at an earlier age. Customary outcome categories used in training algorithmic risk assessments represent adverse contacts with the criminal justice system, such as arrests or convictions. Associations with race are typical here as well.

There are lively, ongoing  discussions about why such racial associations exist (Alpert et al., 2007; Harcourt, 2007; Gelman et al., 2012; Grogger and Ridgeway, 2012; Starr, 2014; Tonrey, 2014; Stewart et al., 2020). Some explanations rest on charges of racial animus in the criminal justice system, some focus on criminal justice practices properly motivated but with untoward effects, some take a step back to larger inequalities in society that propagate crime, and some frame the discussion as but one instance of widespread measurement error caused by sloppy data collection and processing.

It is likely that some mix from each perspective is relevant, but there is no credible integration yet available. Therefore, as a scientific matter we take no position in this paper on explanations for the role of race, although a range of associations are empirically demonstrable. Rather, we more simply build on professed differences in privilege consistent with extensive research (Rothenberg, 2008; Rocque, 2011; Van Cleve and Mayes, 2015; Leonard, 2017; Wallis, 2017; Bhopal, 2018; Edwards et al., 2019; Jackson, 2019; GBD 2019 Police Violence Subnational Collaborators, 2021). From a pragmatic point of view, we are responding as well to common suppositions and frequent stakeholder claims.

We do not require, more generally, that all relevant covariates are included in a criminal justice risk algorithm. An effective forecasting algorithm need not be a causal model for which potential misspecification is always an issue. Indeed, some have argued that an algorithm is not a generative model to begin with (Breiman, 2001; Berk et al, 2022). Although it may at first seem shamelessly expedient, we adopt the practical goal \emph{solely} of more fair and more circumscribed assessments of risk than current practice. Calibration is no longer in the mix. Nor is specification error; with no modeling aspirations, misspecification is not even defined. Empirical evidence evaluating this approach is provided later.

We proceed in three steps: (1) training the risk algorithm in a novel manner, (2) transporting the predictor distribution from the less privileged class to the more privileged class, and (3) constructing conformal prediction sets to serve as risk forecasts. Each step is briefly summarized next.

\subsection{Training The Classifier}

We train a machine learning classifier \textit{only} on White offenders; both the predictors and the outcome are taken solely from Whites. Training on White offenders has several important consequences for our approach to fairness.
\begin{itemize}
\item
No distinctions between White and Black offenders, invidious or otherwise, can be made as the classifier is trained because the algorithm has information exclusively on White offenders; race is a constant. This is not to say that there are no important empirical differences between Black and White offenders. Indeed, there are likely to be important differences that we address shortly with optimal transport.  
\item
Because the training is undertaken solely with data from Whites, claims of measurement error in the training data caused by racial animus or structural racism can be difficult to sustain. For example, a White individual's prior record or post-release chances of arrests are unlikely to be inflated by ``overpolicing.'' Many would argue that a substantial portion of the training data would be tainted by such factors if Black offenders were included.
\item
Training on White offenders provides an algorithmic risk baseline and a potential fairness target for Blacks. Insofar as White offenders are relatively advantaged compared to Black offenders, White algorithmic performance might be seen as desirable for Black offenders; it becomes a target for fairness. One has a prospective answer to the question ``fair compared to what?" The answer is fair compared to how White offenders are treated. No larger fairness claims are made.

\item
White offenders are classified no differently than they would be ordinarily if there were no Black offenders at all. 
\end{itemize}
In short, by training only on Whites, we address many of the fairness complications discussed earlier.

Subsequently, when risk forecasts are sought from unlabeled data for White offenders, one can proceed as usual by obtaining predictions from the trained classifier. Nothing changes from usual practice. However, risk forecasts for unlabeled Black offenders obtained using the White-trained algorithm still can produce racial disparities because Black offenders can have more problematic predictor distributions, whatever their cause. Black offenders on the average will then be treated by the White-trained classifier as particularly risky White offenders, leading to less favorable forecasted outcomes. Another step towards fairness seems to be needed.

\subsection{Transporting Observations Across Protected Groups}

A second adjustment is required to make comparable the joint distribution of predictors for Black offenders and the joint distribution of predictors for White offenders. We use a form of optimal transport (H\"{u}tter and Rigollet, 2020; Manole et al., 2021; Pooladain and Niles-Weed, 2021) to that end. Optimal transport maps a feature vector from the black population to a feature vector from the white population. To take a toy example, an arrested Black offender 18 years of age, with 3 prior robbery arrests and a first arrest at age 14, might be given predictor values from an arrested White offender who was 20 years age with 1 prior robbery arrest, and a first arrest at age 16.

Used in this manner, optimal transport is not a model. Nor is it an exercise in matching. For the toy example, both covariate point masses are within the same $\mathbb{R}^4$. Each observation in the joint covariate distribution for Black offenders is ``transported'' to the closest observation in the joint covariate distribution for the White offenders, where closeness is defined by the squared Wasserstein distance (much like squared Euclidean distance), and the marginal distributions for Black and White offenders are fixed. One has a linear programming formulation in which the sum of the squared Wasserstein distances is made as small as possible subject to the two sets of marginal constraints (Peyr\'e and Cuturi, 2019: chapter 3; H\"utter and Rigollet (2020: section 6.1)). Ideally, the joint covariate \emph{distribution} for Black offenders becomes virtually identical to the joint covariate \emph{distribution} for White offenders, which when used with the White-trained classifier, determines what we mean by treating Black offenders as if they were White offenders when recidivism risks are estimated. 

There is one more technical detail. The transported joint predictor distribution for Black offenders is then smoothed with a form of nonparametric regression so that it can properly be used to transport predictors from new, unlabled cases for which forecasts are needed. Details and pseudocode are provided in Appendix B. Further discussion can also be found in the application.

\subsection{Forecasting for Individual Cases}

The practical task for a criminal justice risk assessment is to forecast one or more behavioral phenomena. By training a classifier only on White offender data and evaluating fairness separately using White test data and transported Black test data, one can as usual compare the aggregate performance of a risk assessment tool across protected classes using conventional confusion tables. There also can be forecasts for individuals that minimize Bayes risk. But we emphasize again that fairness for legally protected classes is being sought. The enterprise remains group fairness. We have nothing to say about whether a given offender is being treated fairly.

Arguably, a more defensible forecasting approach rests on conformal prediction sets (Vovk et al., 2005; 2009; Shafter and Vovk, 2008; Romano et al., 2019; Kuchibhotla and Berk, 2021). The output for a categorical response variable is a prediction set with an associated probability that can include one or more true future outcomes. The prediction set has valid coverage in finite samples of realizations for the population from which the exchangeable data originated. This is consistent with a group approach to fairness. Appendix C provides a more complete treatment and pseudocode for the form of conformal inference we employ. There is further discussion in the application.

\subsection{A Diagrammatic Summary of Our Risk Algorithm}

\begin{figure}[htpb]
\centering
\hspace{-0.25in}\includegraphics[width=1.05\textwidth]{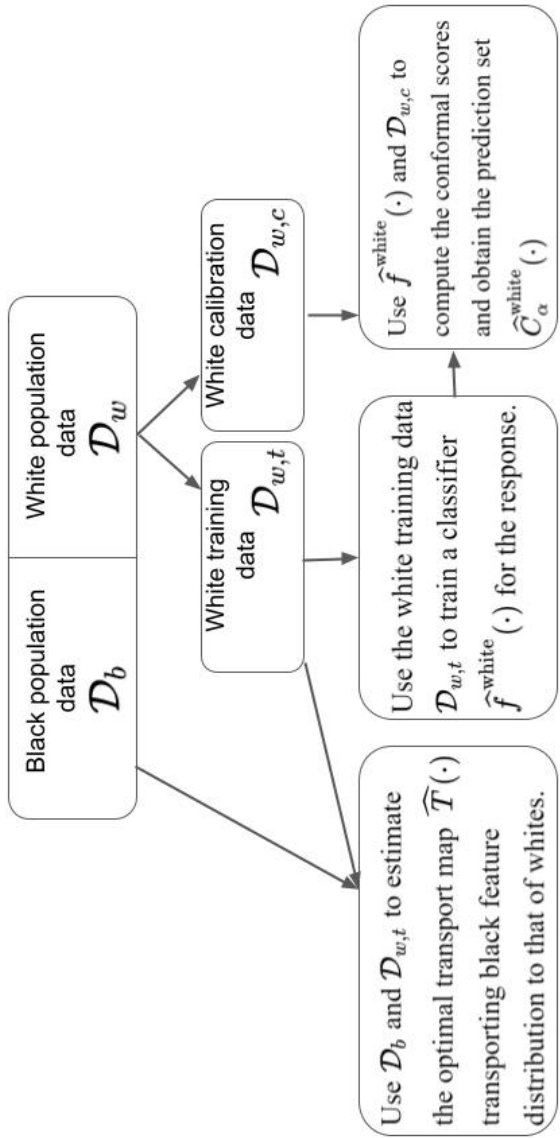}
\caption{Flowchart for the training part of our ``fair'' risk algorithm. The function $\widehat{f}^{\texttt{white}}(\cdot)$ denotes the output vector of probabilities for each outcome obtained from a classifier fit on the White training data. The map $\widehat{T}(\cdot)$ denotes the estimate of the optimal transport map; see Appendix~\ref{appsec:optimal-transport} (and Algorithm~\ref{alg:optimal-transport-map}) for details. The set $\widehat{C}_{\alpha}^{\texttt{white}}(\cdot)$ is the conformal prediction set obtained using $\widehat{f}^{\texttt{white}}(\cdot)$ and the White calibration data; see Appendix~\ref{appsec:conformal-inference} (and Algorithm~\ref{alg:nested-conformal-prediction}) for details.}
\label{fig:flowchart1}
\end{figure}

Procedural details are provided in two diagrams with brief explanations for our entire risk algorithm from the training of a classifier, to the use of optimal transport, to the output of a conformal prediction set. Figure~\ref{fig:flowchart1} addresses how the fair algorithm was constructed. Figure~\ref{fig:flowchart2} addresses fair forecasting. Both figures are further unpacked in the appendices. In addition, the data analysis to come provides a grounded methodological discussion.

\begin{figure}[htpb]
\centering
\includegraphics[width=4.5in]{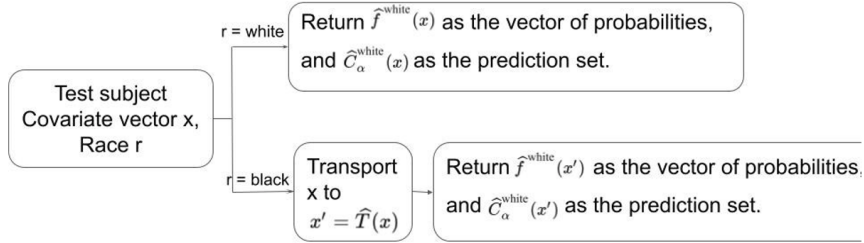}
\caption{Flowchart for producing fair conformal prediction sets. The function $\widehat{f}^{\texttt{white}}(\cdot)$ denotes the output vector of probabilities for each outcome obtained from a classifier fit on the first split White training data. The map $\widehat{T}(\cdot)$ denotes the estimate of the optimal transport map obtained from the first split training data for Whites and Blacks; see Appendix B for details. The set $\widehat{C}_{\alpha}^{\texttt{white}}(\cdot)$ is the conformal prediction set obtained using $\widehat{f}^{\texttt{white}}(\cdot)$ and second split white training data.}
\label{fig:flowchart2}
\end{figure}

In theory, we achieve our overall goal of constructing a fair algorithmic risk tool. But in practice, there are two challenges. First, our procedures must be implemented with data having many potentially important offenders features discarded, usually for legal or political objections. Second, as a formal matter, external fairness parities cannot be evaluated from test data without very strong assumptions discussed in Appendix D. Recall that the required counterfactual outcome of a fair arrest generating process is unobservable for Black offenders in existing criminal justice datasets. 

\subsection{A Pareto Improvement for Protected Groups}

Our procedures to improve fairness arguably have a more general justification. Specifically, we claim that at the level of legally protected classes, there is a Pareto improvement. Consistent with our focus on group fairness, no formal claims are made about the consequences for individuals. But in practice, Blacks may be made better off and Whites may not be made worse off on the average.

We are \emph{provisionally} proceeding as if White offenders at arraignment are members of an advantaged class compared to Black offenders at arraignment. There is no claim that all individual White offenders are privileged compared to all individual Black offenders. The issues can subtle, as legal scholars have emphasized (Hellman, 2020: 449 -- 552). 

By training only on White offenders and transporting the Black covariate distribution to the White covariate distribution, the presence of Black offenders has absolutely no impact on risk assessments for Whites. Nothing changes, and the \emph{class} of Whites is not made worse off. But the risk algorithm treats all Black offenders as if they are White. If the class of Whites is really advantaged at arraignment, the \emph{class} of Black offenders can be made better off, or at least are not made worse off. In principle, this is a Pareto improvement for legally protected classes. We will bring data to bear later.

To underscore the difference between legally protected classes and class members, consider the Civil Rights Act of 1964 as amended in 1972 to prohibit sex discrimination in educational institutions receiving federal financial assistance (i.e, ``Title IX''). The landscape of college sports changed dramatically such that women could expect to have their own intercollegiate sports teams. As a class, women benefited, even if the direct benefits were conferred largely on the small fraction of all female students who became varsity athletes. If athletic department budgets were increased sufficiently, a Pareto improvement was possible, determined together by the class of men and the class of women.
 
\section{The Data} 

We analyze a random sample of 300,000 offenders at their arraignment from a particular urban jurisdiction in the United States.\footnote
{
At an arraignment, which is supposed to be held within 48 hours of an arrest, the charges are read officially to the arrested offender. The presiding magistrate then decides whether the offender can be released, sometimes with a bail bond, subject to a later return to court, or detained until that later court date.
} 
Because of the random sampling, the data can be treated as IID and, therefore, exchangeable. Even without random sampling, one might well be able to make an IID case because the vast majority of offenders at their arraignment are realized independently of one another.

Among those being considered for release at their arraignment, one outcome class (coded ``1'') to be forecasted is whether the offender will be arrested after release for a \emph{crime of violence}. An absence of such an arrest (coded ``0'') is the alternative outcome class to be forecasted. Violent crimes usually are of particular concern.\footnote
{ 
It might seem that using a conviction rather than an arrest would convey more about the actual crime, but the vast majority of criminal trials are resolved by a guilty plea after a negotiated agreement between the defense and prosecuting attorneys. Strategic maneuvering can dominate the process. Racial factors can enter as well. 
} 

The follow-up time was 21 months after release. For reasons related to the ways in which competing risks were defined, 21 months was chosen as the midpoint between 18 months and 24 months. For the analysis to follow, the details are unimportant.

Candidate predictors were the usual variables routinely available in large jurisdictions. Many were extracted from adult rap sheets and analogous juvenile records. Biographical variables included race, age, gender, residential zip code, employment information, and marital status. There were overall 70 potential predictors. 

In response to potential stakeholder concerns about ``bias,'' we excluded from the classifier training race, zip code, marital status, employment history, juvenile record, and arrests for misdemeanors and other minor offenses. Race was excluded from the training for obvious reasons. Zip code was excluded because, given residential patterns, it could be a close proxy for race.  ``Close'' was not clearly defined, but if offenders' zip codes were known, one usually could be quite sure about their race. No other potential predictors were seen by stakeholders as close proxies for race.\footnote
{
Even if zip code were included as a predictor, racial differences would not have been built into the training. Only the zip codes of White offenders would have been used. The same applies to any race proxies. Moreover, distributional differences by race would have been removed subsequently when optimal transport was applied. Indeed, that is the point of applying optimal transport.
}
Employment history and marital status were eliminated because there were objections to using ``life style'' measures. Juvenile records was discarded because poor judgement and impulsiveness, often characteristics of young adults, are not necessarily indicators of long term criminal activity. Minor crimes and misdemeanors were dropped because many stakeholders believed that arrests for such crimes could be substantially influenced by police discretion, perhaps motivated by racial animus. 

The truth underlying such concerns is not definitively known, but insofar as the discarded predictors were associated with any included predictors, potential biases remain (Berk, 2009). These decisions underscore our earlier point that there are legitimate disagreements over what features of individuals should determine when a similarly situated comparison has been properly undertaken. They also highlight the trade-offs to be made when a suspect predictor also is an effective predictor.

In the end, the majority of the predictors were the number of prior arrests separately for a wide variety of serious crimes, and the number of counts for various charges at the arraignment. Other included predictors were whether an individual was currently on probation or parole, age, gender, the age of a first charge as an adult, and whether there were earlier arrests in the same year as the current (arraignment) arrest. For the analyses to follow, there were 21 predictors.\footnote
{
The two age-related variables and whether there were other arrests within the past year are ``dynamic variables'' because they can change over time. For other criminal justice decisions, such as whether to grant parole, there can be many more dynamic variables (e.g., work history in prison). At an arraignment, one is limited largely to what could be extracted from existing rap sheets and current charges.
}

The 300,000 cases were randomly split into training data for White offenders, training data for Black offenders, test data for White offenders, and test data for Black offenders. Half the dataset was used as training data ($n = 150,000$) and half the dataset was used as test data ($n = 150,000$). Sizes of the racial splits of the training and test data were simply determined by the numbers of Black offenders and White offenders available in the data. Each racial split had at least 40,000 observations.

\section{Fairness Results} 

We began by training a stochastic gradient boosting algorithm (Friedman, 2001) on White offenders only using the procedure \textit{gbm} from the library \textit{gbm} in the scripting language \textit{R}. For illustrative purposes and consistent with many stakeholder priorities, the target cost ratio was set at 8 to 1 (Berk, 2018). Failing to correctly classify an offender who after release is arrested for a crime of violence was taken to be 8 times worse than failing to correctly classify an offender who after release is not arrested for such a crime. We were able to approximate the target cost ratio reasonably well in empirical confusion tables by weighting more heavily training cases in which there was an arrest for a crime of violence. This, in effect, changes of the prior distribution of the outcome variable. 

All tuning defaults worked satisfactorily except that we chose to construct somewhat more complex fitted values than the defaults allowed.\footnote
{
We used greater interaction depth to better approximate interpolating classifiers (Wyner et al., 2015). Even after weighting, we were trying to fit relatively rare outcomes. We needed an ensemble of regression trees each with many recursive partitions of the data.
}
The results were essentially the same when the defaults were changed by modest amounts. The number of iterations (i.e. regression trees) was determined empirically when, for a binomial loss, the reductions in the test data effectively ceased.\footnote
{ 
Because of the random sampling used by the \textit{gbm} algorithm, the number of iterations in principle can vary a bit with each fit of the data. Also, the number of trees can arbitrarily vary by about  25\% with very little impact. 
} 
 
\subsection{Algorithmic Performance Results for White Offenders }

Algorithmic risk assessments can be especially challenging when the marginal distribution of the outcome is highly unbalanced. For binary outcomes, this means that if criminal justice decision-makers always forecast the most common outcome class, they will be correct the vast majority of the time. It is difficult for an algorithm to forecast more accurately. Because post-arraignment arrests for a crime of violence are well-known to be relatively rare, we were faced with the same challenge that, nevertheless, provided an instructive test bed for examining fairness.

To set the stage, Table~\ref{tab:t1} is the confusion table for White offenders using the risk algorithm trained on Whites and test data for Whites.\footnote
{
The empirical cost ratio in Table~\ref{tab:t1} is 11246/1527, which is 7.4 to 1. It is very difficult in practice to the arrive exactly at the target cost ratio with test data, but cost ratios within about 20\% of the target usually lead similar confusion tables.
}
Perhaps the main message is that if arraignment releases were precluded solely by the risk algorithm when arrests for a violent crime were forecasted, more violent crime might be prevented. 

Here is the reasoning. From the outcome marginal distribution of an arrest for a crime of violence, minimizing Bayes loss always counsels forecasting no such arrest after an arraignment release. That forecast would be wrong for 7.5\% of the White offenders. From the left column in Table~\ref{tab:t1}, when the algorithm forecasts no arrest for a violent crime after an arraignment, the forecast is wrong for 5\% of the White offenders. If from the marginal distribution one always forecasted an arrest for a crime of violence, the forecast would be wrong for 92.5\% of the White offenders. From right column in Table~\ref{tab:t1}, the algorithm is mistaken for 85\% of the white offenders. These are modest improvements in percentage units, but given the large number of White offenders, over 2000 of violent crimes might be averted if the risk algorithm determined the arraignment release decision rather than the presiding magistrate. By this standard, the risk algorithm performs better.

\begin{table}[h]
\scriptsize
\caption{Test Data Confusion Table for White Offenders Using the White-
Trained Algorithm (28\% Predicted to Fail, 7.5\% Actually Fail)}
\begin{center}
\begin{tabular}{|c|c|c|c|}
\hline 
 Actual Outcome & No Violence Predicted &Violence Predicted & Classification Error  \\ 
 \hline 
 No Violence & 31630 (true positives) & 11246 (false positives) & .26 \\
 Violence & 1527 (false negatives) & 1975 (true negatives) &  .47 \\
 \hline
 Forecasting Error & .05 & .85 &  \\
 \hline \hline
\end{tabular}
\end{center}
\label{tab:t1}
\end{table}

Yet, forecasting accuracy is shaped substantially by the cost ratio. Because the target cost ratio treats false negatives as 8 times more costly than false positives, predictions of violence in Table~\ref{tab:t1} are dominated by false positives. This follows directly and necessarily from the imposed trade-offs. Releasing violent offenders is seen to be so costly that even a hint of future violence is taken seriously. But then, many mistakes are made when an arrest for a crime of violence is forecasted. In trade, when the algorithm forecasts no arrest for a violent crime, it is rarely wrong; there are relatively few false negatives. This too follows from the target cost ratio. If even a hint of violence is taken seriously, those for whom there is no such hint are likely to be very low risk releases. In short, with different target cost ratios, the balance of false positives to false negatives would change, perhaps dramatically, which means that forecast accuracy would change as well.\footnote
{
Keep in mind that ``An algorithm is nothing more than a very precisely specified series of instructions for performing some concrete task''(Kearns and Roth, 2020: page 4). It is not meant to explain some phenomenon, depict causal effects, or characterize how the data were generated. Consequently, the forecasted outcomes have nothing to say about \emph{why} either outcome is realized. Some arrests might be ``rightous,'' some might be a direct or indirect product of race, and many can be a mix of the two, but the precise mechanisms are not manifest. 
}

The aversion to false negatives contributes to a projection that 28\% of the White offenders will fail through a post-release arrest for a violent crime. In the test data, only 7.5\% actually fail in this manner. The policy-determined trade-off between false positives and false negatives produces what some call ``overprediction." With different trade-off choices, overprediction could be made better or worse. In either case, there would likely be important concerns to reconsider.

Overprediction worries become even more prominent if test data for Black offenders are used to forecast post-arraignment crime. When the Black test data are employed with the algorithm trained on Whites, \emph{41\% of the Black offenders are forecasted to be arrested for a crime of violence, whereas 11\% actually are.} The re-arrest base rate is a bit higher for Black offenders than White offenders (i.e., 11\% compared to 7.5\%), but the fraction projected to be arrested for a violent crime increases substantially: \emph{from 28\% to 41\%}. As emphasized earlier, the latter disparity cannot be a product of racial differences in the algorithmic machinery. It is trained only on Whites. The likely culprit is racial disparities in the feature distribution provided to the classifier. In any case, there is clear evidence from the test data that prediction parity is not achieved solely by training the risk algorithm on White offenders. 

Table~\ref{tab:t1} also provides for White offenders conventional test data performance statistics for the false positive rate, the false negative rate, forecasting accuracy for an arrest for a crime of violence, forecasting accuracy for no arrest, and the empirical cost ratio. For example, the false positive rate is .26 and when no arrest is forecasted, it is wrong 5\% of the time. Comparisons could be made to the full confusion table for Black offenders, but the limitations of internal fairness would intrude; recall that historical test data are poorly suited for examinations of external fairness. We postpone further discussion until more results are reported. 

Overall, performance is roughly comparable to other criminal justice risk assessments and probably worth close consideration by stakeholders (Berk, 2018). No doubt some changes in the risk classifier would be requested, and the results would be reviewed for potential alternatives to implement. Our intent, however, is not to claim that the results in Table~\ref{tab:t1} are definitive. Rather, they provide a realistic context for an empirical consideration of fairness. Clearly, prediction parity is at this point problematic.

\subsection{Optimal Transport Performance}

We applied optimal transport, briefly described earlier, using the procedure \textit{transport} in R.\footnote
{
There are several computational options for estimating the coupling matrix. We used the default ``revsimplex'' (Luenberger and Ye 2008, Section 6.4) that worked very well.
}
No tuning in a conventional sense was needed. However the coupling matrix was $n \times n$ which meant that memory considerations came to the fore. We tried 1000, 2000, 3000, 4000 and 5000 observations in ascending order. At 5000 observations, computer memory was exceeded. We proceeded, therefore, using 4000 randomly selected test data observations for Black offenders.\footnote
 {
 We were surprised by how well \textit{transport} performed with just 4000 observations. At first, we were skeptical, and we tried several toy examples and datasets previously used by others. We found no reasons to discount our results.
 }
 
A key diagnostic for optimal transport in practice is how well the transported source joint probability distribution compares to the destination joint probability distribution. Summary fit statistics are too coarse. They can mask more than they reveal. A better option is compare the correlation matrices from the two distributions. For these results, there were no glaring inconsistencies, but it was difficult to translate differences in the correlations into implications for fairness. 

An instructive diagnostic simply is to compare the marginal distributions for each predictor. Using histograms, we undertook such comparisons for each of the 21 predictors. The following figures show the results for the predictors that dominated the fit when the risk algorithm was trained on the data for White offenders; these are the predictors that mattered most. There were similar optimal transport results for the other, less important, predictors.
 
\begin{figure}[htp]
\centerline {
\includegraphics[scale=.45]{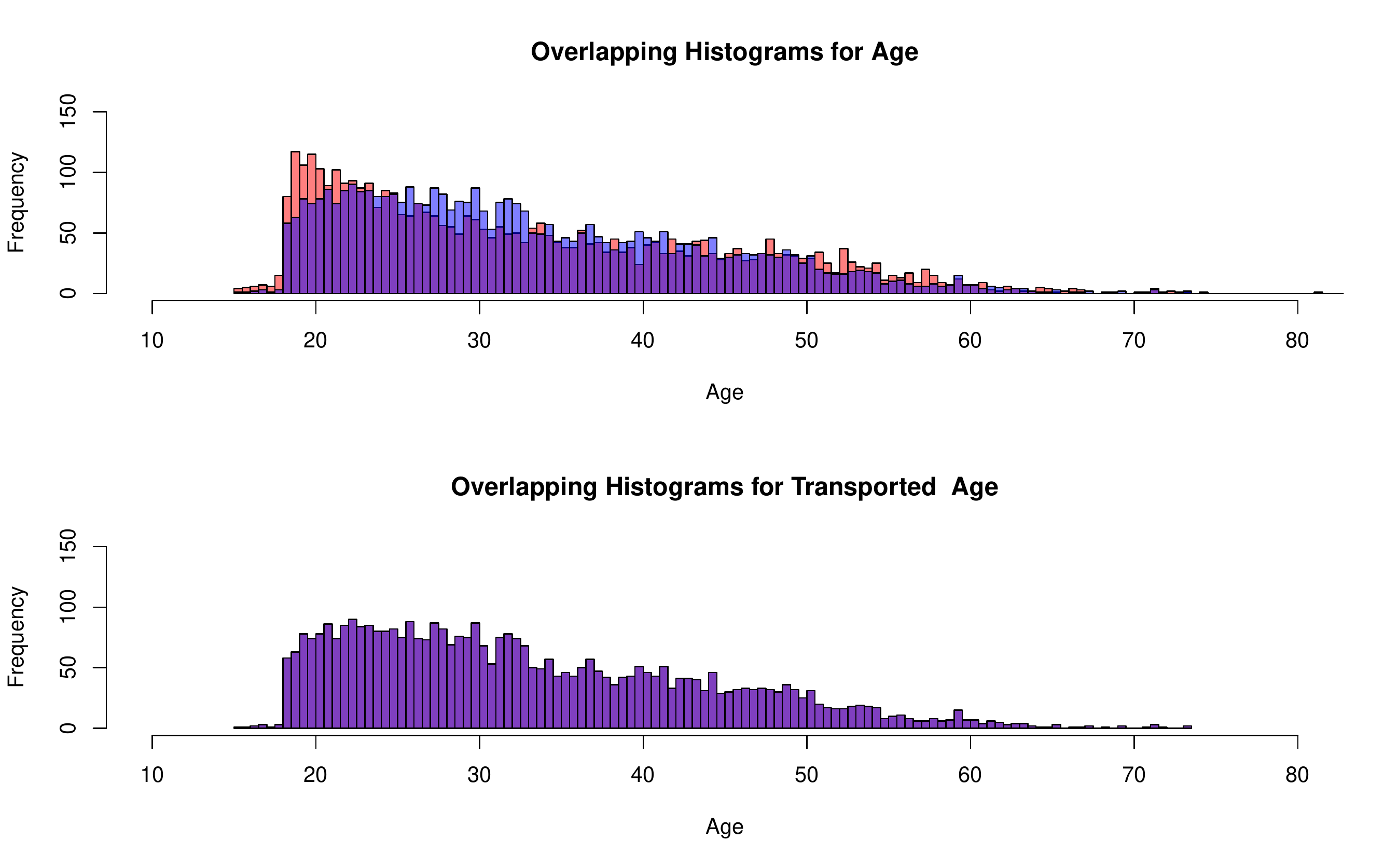}
}
\caption{Histograms for an Offender's Age and Transported Age (Black offenders in orange, White offenders in blue, overlap in purple, N= 4000)}
\label{fig:hist1}
\end{figure}

It is well-known that younger individuals have a greater affinity for violent crime than older individuals. Figure~\ref{fig:hist1}, constructed from the 4000 randomly selected observations provided to the procedure \textit{transport}, shows the results for the age of the offender. The top histogram compares the test data distribution for Whites in blue to the test data distribution for Blacks in orange. The purple rectangles show where the two distributions overlap. Clearly, Black offenders at arraignment are on the average somewhat younger, especially for the youngest ages that place an offender at the greatest risk. The bottom histogram is constructed in the same manner but now, the White age distribution from test data is compared to the transported Black distribution. There are no apparent differences between the two. Clearly, Black offenders are no longer over-represented among the youngest ages.

One must be clear that Figure~\ref{fig:hist1} shows how the age \emph{distribution} for the White test data and the Black transported test data are made virtually indistinguishable. We emphasize that this does not imply exact one to one matching of Black offenders to White offenders in units of years. Figure~\ref{fig:hist1} results from a linear programming solution in which, as noted earlier, the squared distances between observations from the two joint covariate distributions in predictor space are minimized subject to fixed marginal distributions. More details are provided in Appendix C. 

The performance of optimal transport in the bottom histogram may seem too good to be true. However, despite some distributional differences in the top histogram that may matter for risk, the overall shape of the two distributions is very similar. Both peak at low values and gradually decline in an almost linear fashion toward a long right tail. One should expect optimal transport to perform well under such circumstances.

Perhaps more surprising is that the two distributions are so similar to begin with. But arrests are a winnowing process that can affect members of protected classes in similar ways. The pool of individuals who are arrested is more alike than the overall populations from which they come. Regardless of race, the pool disproportionately tends to be young, male, unemployed, and unmarried, with appreciable previous police contact. It is commonly said that less than 10\% of the overall population are responsible for more than 50\% of the crime (e.g., Nath, 2006) This disparity is reflected in the backgrounds of individuals who are arrested, coming more likely from that 10\%.

\begin{figure}[htp]
\centerline {
\includegraphics[scale=.45]{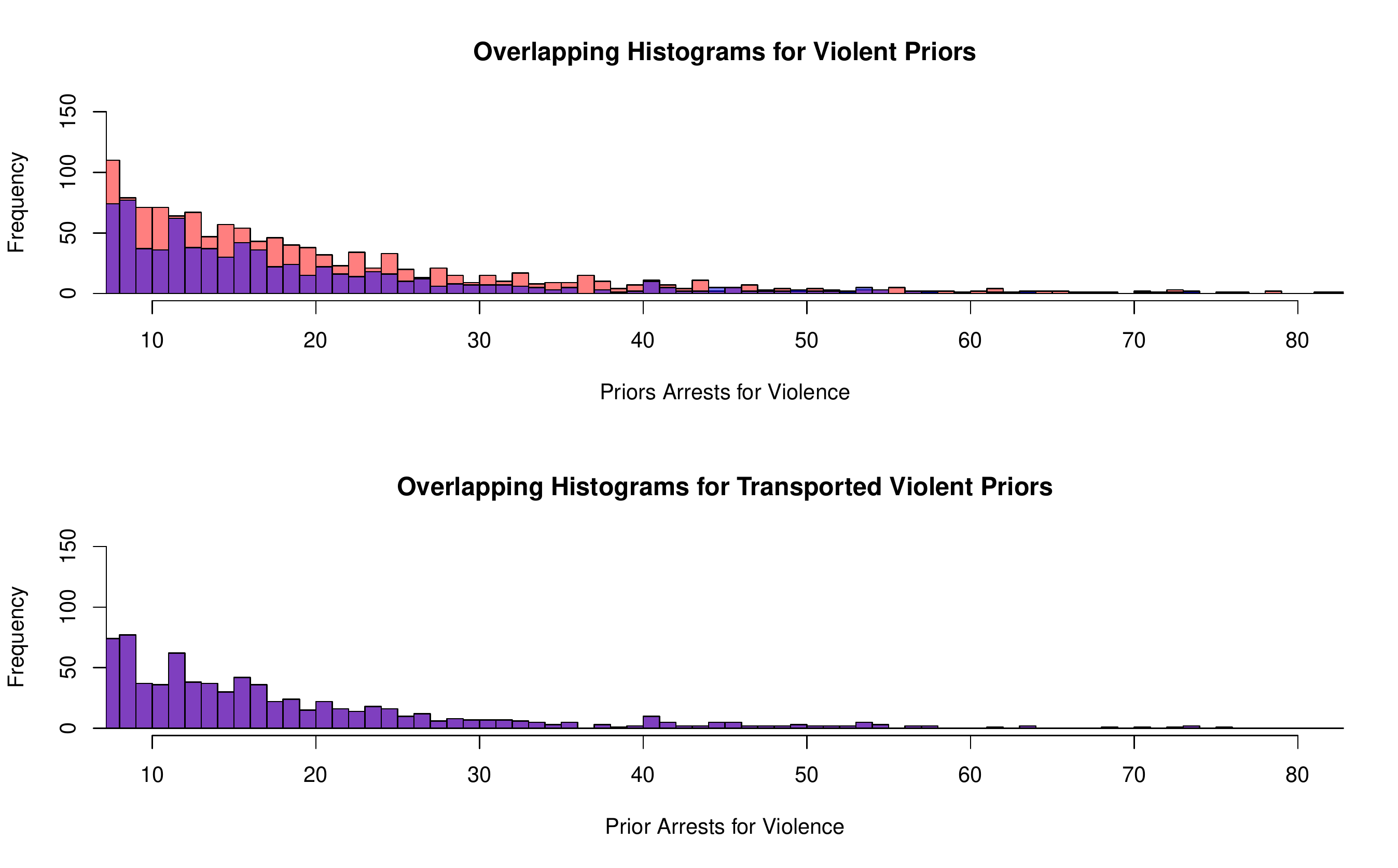}
}
\caption{Histograms for the Number of Prior Arrests for a Crime of Violence and the Transported Number of Prior Arrests for a Crime of Violence (Black offenders in orange, White offenders in blue, overlap in purple, N= 4000)}
\label{fig:hist2}
\end{figure}

Figure~\ref{fig:hist2}, constructed from the same 4000 observations, shows the results for an offender's number of prior arrests for crimes of violence, which is also known to be associated with post-arraignment violent crime. It is apparent in the top histogram that Black offenders have more priors up to about 40, at which point there are too few cases to draw any conclusions. After the application of optimal transport, the bottom histogram shows no apparent differences. As before, the two distributions were not dramatically different before optimal transport was applied.

\begin{figure}[htp]
\centerline {
\includegraphics[scale=.43]{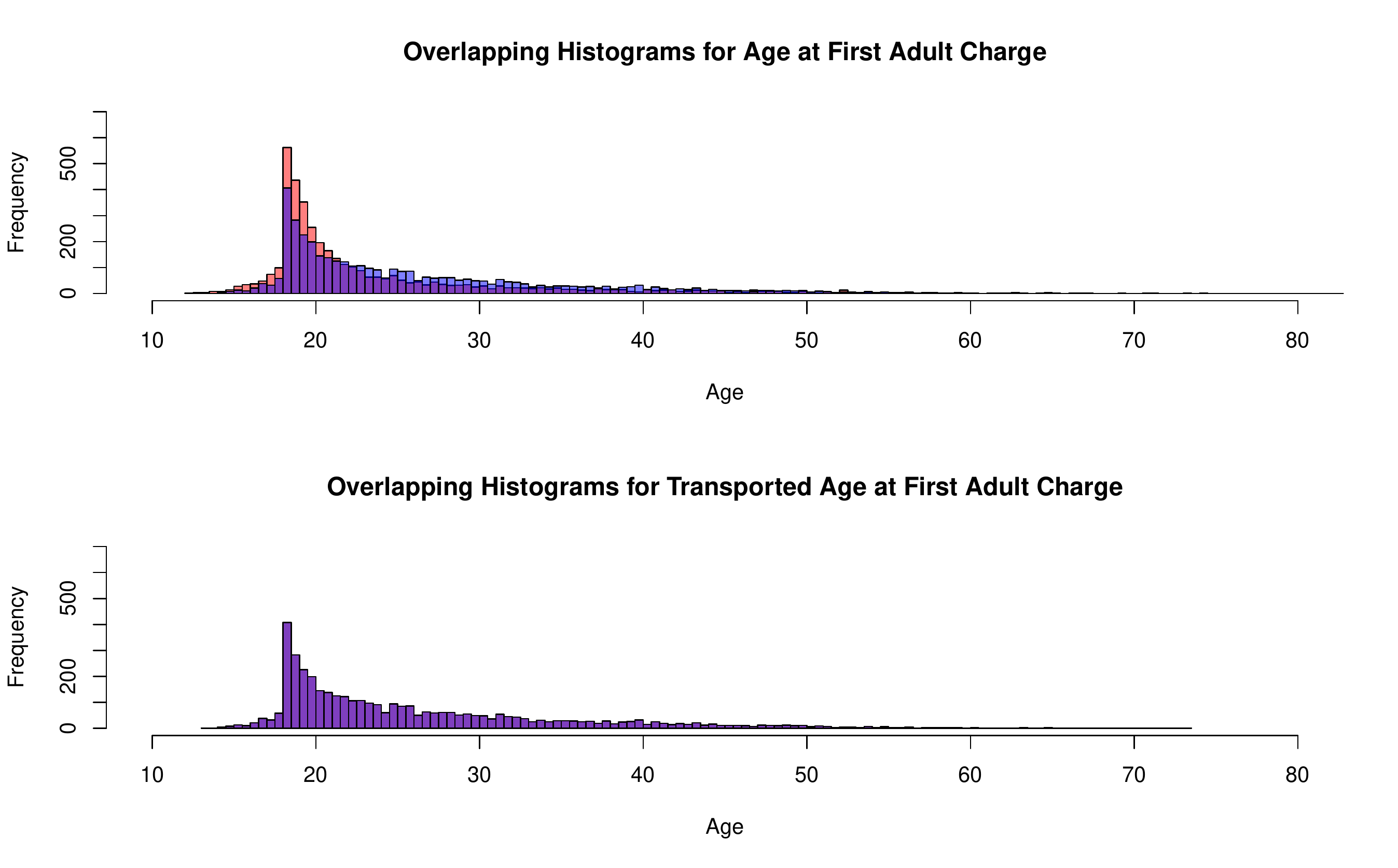}
}
\caption{Histograms for the Age of the first Adult Charge and the Transported Age of the first Adult Charge (Black offenders in orange, White offenders in blue, overlap in purple, N = 4000)}
\label{fig:hist3}
\end{figure}

Figure~\ref{fig:hist3}, using the same 4000 observations, shows the results for the earliest age at which an offender was charged as an adult. Offenders who start their criminal activities at a younger age are more crime-prone subsequently. From the top histogram, Black offenders are more common than White offenders at the younger ages. That disparity disappears in the bottom histogram after optimal transport is applied, no doubt aided by the similar shapes of the two distributions. Optimal transport seems to perform as hoped to remove racial disparities in the two joint predictor distributions. 

But, the effectiveness of optimal transport in a \textit{forecasting} setting remains to be addressed. We converted the transported joint predictor distribution constructed from the Black offender test data into conformal scores like those used in forecasting. The classifier trained on White offenders was tasked with producing the probabilities of an arrest for a violent crime. These probabilities were then subtracted from ``1'' and from ``0,'' yielding Black offender conformal scores for the two possible outcome classes. In other words, we were proceeding for illustrative purposes as if the Black test data were unlabeled, just as new data would be when forecasts are needed. Conformal prediction procedures for more than two outcome classes can be found in Kuchibhotla and Berk (2021).

\begin{figure}[htp]
\centerline {
\includegraphics[scale=.40]{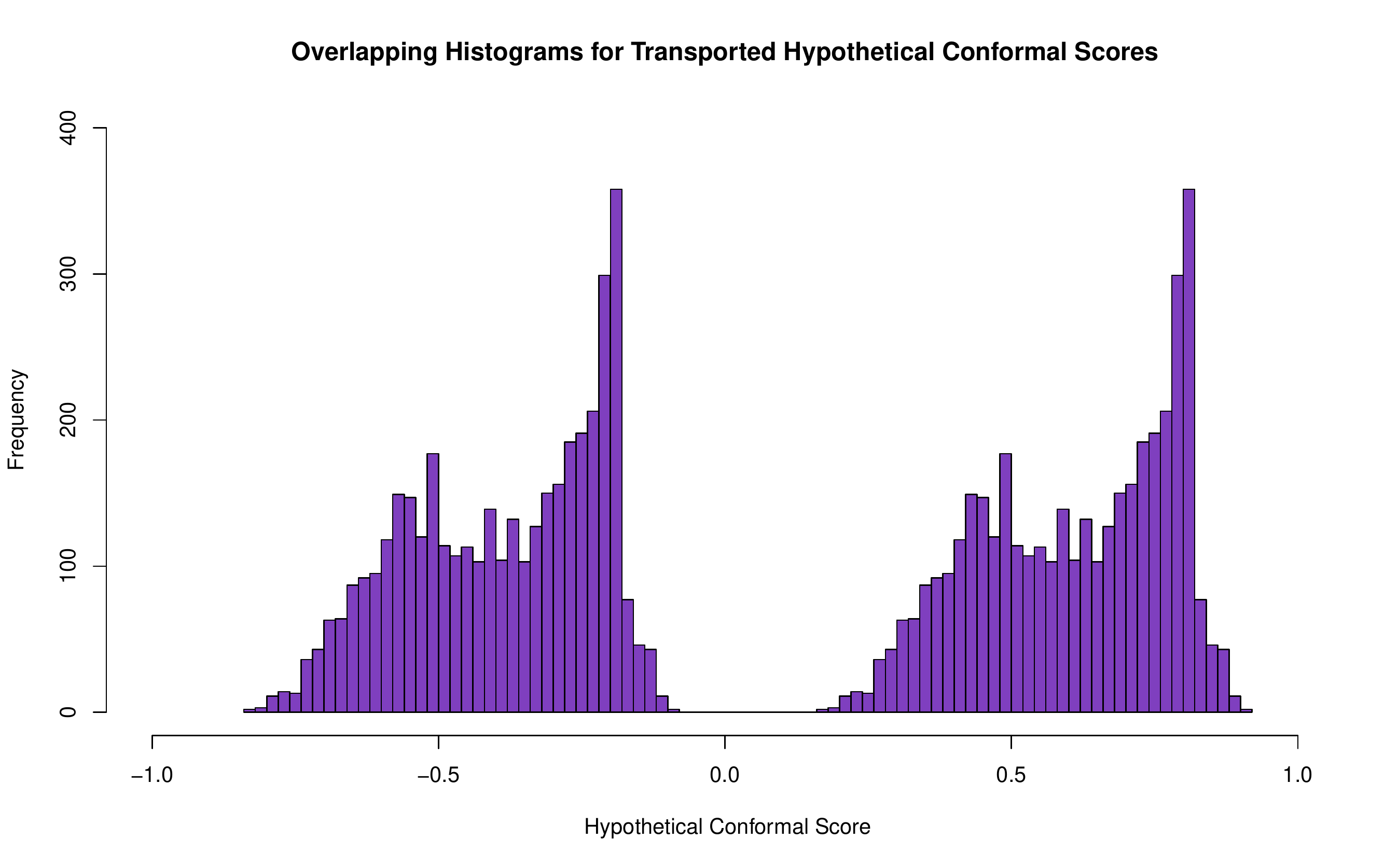}
}
\caption{Histograms for White Conformal Scores and Transported Black Conformal Scores (Black offenders in orange, White offenders in blue, overlap in purple, N = 4000)}
\label{fig:hist4}
\end{figure}

The two conformal score distributions, one for the forecasted 1s and one forecasted 0s, were then compared to the White offender conformal scores computed in the same manner from the White test data (i.e., as if the data were unlabeled). Ideally, there would be no apparent racial differences. 

Figure~\ref{fig:hist4} shows the results. The histogram to the left contains the conformal scores for cases in which the hypothetical outcome is no arrest for a crime of violence.  The histogram to the right contains the conformal scores for cases in which the hypothetical outcome is an arrest for a crime of violence. As before, the histogram rectangles for Black offenders are in orange, the histogram rectangles for White offenders are in blue, and the overlap rectangles are in purple. Both histograms are entirely purple. The test data distribution of conformal scores for White offenders and Black offenders are for all practical purposes the same.\footnote
{
The N for Whites set a 4000 because that is the number of cases used by the optimal transport procedure. This also makes comparisons between histograms easier to implement.
}
The claim is strengthened that for classifiers trained on White data, conformal prediction parity might be improved by optimal transport.

When actual forecasts are required, there is another step. For Black offenders, one needs a procedure that converts the predictor values for each unlabeled case into its corresponding transported values. These new cases were not available for the earlier optimal transport exercise, and repeating optimal transport for each new unlabled case was at least impractical. H\"utter and Rigollet (2020), instead suggest fitting a multivariate nonparametric smoother and using that to get good approximations of transported conformal scores. An added benefit is that the full range of predictor values for the unlabeled data can have comparable transported values. We applied random forests.\footnote
{
For example, an age of 57 for an unlabled case may not exist in the transported data unless a smoother is applied. 
Random forests solves such problems because the inequalities responsible for recursive partitioning tree by tree leave no gaps in predictor values.
}

One begins with a conventional $n \times p$ matrix of the original joint predictor test data distribution for Black offenders denoted by $X^{Test}_{b}$. One also has an $n \times p$ matrix of the transported joint predictor distribution for Black offenders denoted by $X^{Trans}_{b}$. Each column of  $X^{Trans}_{b}$ is regressed in turn on $X^{Test}_{b}$. Here, that means repeating this operation 21 times. Subsequently, the predictors for any unlabeled case could be used as input for the fitted random forest to output approximations of each transported predictor. These are then collected in a matrix denoted by $\hat{X}^{Trans}_{b}$. When conformal scores for forecasting are needed, these approximations can be employed as usual as if they were the actual transported predictor values. 

\begin{figure}[htp]
\centerline {
\includegraphics[scale=.40]{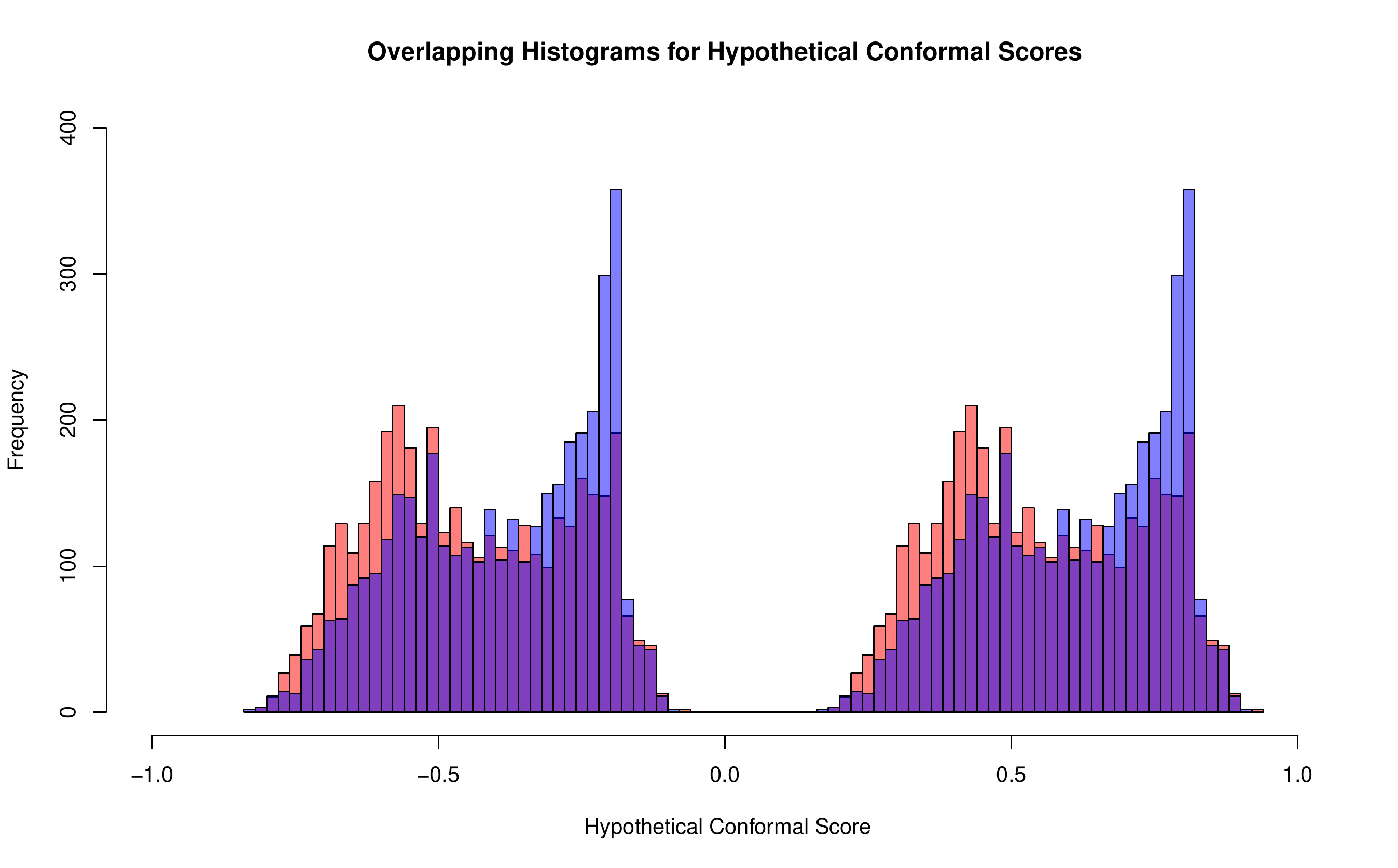}
}
\caption{Histograms for White Conformal Scores and Smoothed Black Conformal Scores with the 0 Outcome Label on the Left and 1 Outcome Label on the Right (Black offenders in orange, White offenders in blue, overlap in purple, N= 4000)}
\label{fig:hist5}
\end{figure}

There is evidence from Figure~\ref{fig:hist5} that some of the overlap in Figure~\ref{fig:hist4} is lost because of the random forest approximation. For both histograms, Black offenders are somewhat overrepresented at smaller values and White offenders are somewhat overrepresented at larger values. The performance of optimal transport has been degraded. If a conventional prediction region were imposed, Black offenders might be more commonly forecasted to be arrested for a violent crime and White offenders might be more commonly forecasted not to be arrested for a violent crime. 

The results might improve if our random forests application were better tuned or if some superior fitting procedure were applied. But, the conformal scores that matter for the contents of conformal prediction sets are only those that fall in the near neighborhood of the prediction region's boundaries. Some may view this as a form of robustness. We need to consider whether in practice the reduction is overlap matters for fairness.

\section{Evaluating Fairness in the Algorithmic Determinations of Risk}

Recall that even after training the classifier only on White offenders, racial disparities remained. Especially relevant for this analysis the lack of prediction parity. The apparent unfairness is caused by differences at arraignment between the joint predictor distributions for Black offenders and White offenders. We have shown that optimal transport can remove such disparities. But they are perhaps re-introduced when forecasts need to be made.

We have argued elsewhere (Kuchibhotla and Berk, 2021) that when forecasting is the goal, accuracy is more usefully captured by conformal prediction sets than by confusion tables. One major problem with confusion tables is that forecasts are justified by minimizing Bayes loss even if there are very small differences between the estimated probabilities. For example, a distinction of .90 versus .10 for an arrest compared to no arrest, produces same forecasted outcome category as a distinction of .51 versus .49. In other words, the \textit{reliability} of the forecast is ignored, and an algorithm is forced to make a decision when perhaps it would be better to ``defer'' (Madras et al. 2018).  Another problem is that there are no finite sample coverage guarantees for confusion table forecasts, which can be problematic for real decisions when the number of cases is modest. 

Nevertheless, standard practice and many scholarly treatments of fairness emphasize examinations of confusion tables. We take a rather different approach using conformal prediction sets. But interested readers can see in Appendix E that if one transports for Black offenders their joint distribution including the \emph{response variable} as well the predictors, the confusion table from test data for White offenders is effectively the same as the confusion table from transported test data for Black offenders. For many stakeholders and some scholars, this may suffice for algorithmic fairness. This result should be produced in general. However, any conclusions are based on historical data which, as we have discussed, can be misleading for most kinds of fairness claims. 

\subsection{Results for Conformal Prediction Sets}

We remain centered on prediction parity. For these data, predictive parity requires that conformal prediction sets for Black offenders and White offenders are substantially the same for a given coverage probability. Table~\ref{tab:t2} shows the results for our data on offenders at arraignment. A coverage probability was a specified somewhat arbitrarily as $.95$.\footnote
{
The coverage probability is, in effect, a tuning parameter that can be varied by the researcher. For example, with smaller coverage probabilities, the conformal prediction sets tend to contain fewer elements. There are potential gains in precision in trade for losses in certainty.
}

\begin{table}[htp]
\begin{center}
\begin{tabular}{|c|c|c|c|}
\hline \hline 
Prediction & White  & Black & Black\\
Set           & Test Data & Transported data & Smoothed data \\
\hline
$\{\emptyset\}$  & 0.0  & 0.0 & 0.0 \\
$\{0\}$ &  0.58  & 0.58  & 0.54\\
$\{1\}$ & 0.03    & 0.03 & 0.03\\
$\{1,0\}$ & 0.39  & 0.39  &0.43\\
 \hline \hline
\end{tabular}
\end{center}
\caption{Estimated Proportions for Conformal Prediction Sets from White Test Data Predictors, Transported Black Predictors and Fitted Transported Black Predictors all at $1-\alpha=.95$}
\label{tab:t3}
\end{table}

In order to obtain a sufficient number of observations for instructive results, we treated the test data for White and Black offenders as if the labels were unknown. For White offenders, we obtained conformal prediction sets as one ordinarily would. This is a very important feature of our procedures because it guarantees that our procedures do not alter the conformal prediction sets computed for Whites. Overall, therefore, White offenders as a protected class are not made worse off by the fairness adjustments because there are no such adjustments for White offenders.

For Black offenders, we proceeded in the same manner except using two different predictor distributions: for the transported, joint predictor distribution and for its random forests smoothed, transported approximation. For Black offenders, there were 4000 observations. For White offenders, there about 10 times more.

The rows in Table~\ref{tab:t2} contain results for the four possible conformal prediction sets, where ``1'' denotes an arrest for a violent crime and ``0'' denotes no such arrest. These prediction sets are shown in the first column on the left. In the second column are the proportions of times each prediction set materialized for the White test data. There were no empty sets implying that there were no outlier conformal scores.  The most common prediction set was $\{0\}$ followed closely by $\{1,0\}$. The prediction set $\{1\}$ surfaced very rarely. One might have expected these results because the outcome class of no arrest for a crime of violence dominated the marginal distribution of the response variable. 

One important implication from the conformal prediction sets for Whites is that a substantial number of the arrest forecasts produced by the risk classifier and reported Table~\ref{tab:t1} might properly be seen as unreliable. When an arrest forecasted, the boosting classifier very often was unable produce a definitive distinction between the two possible outcome classes. Yet, unreliable forecasts are treated by confusion tables the same as reliable forecasts.

The second and third columns have identical prediction set proportions for up to two decimal places; the prediction sets from the White offenders' test data and from the Black offenders'  transported test data are effectively identical. One has prediction parity in principle. 

The fourth column shows that in practice there also is very close to perfect prediction parity when the approximate transported predictors are used for Black offenders. The proportion of prediction sets for which the forecast is an arrest for a violent crime remains at .03 for both Blacks and Whites. There is a slight reduction in the proportion of prediction sets that include no arrest by itself (i.e., $\{0\}$) and a slight increase in the proportion of prediction sets for which the classifier cannot make a clear choice (i.e., $\{1,0\}$). Whether such differences matter would be for stakeholders to decide. A lot would depend on what a presiding magistrate would do with the $\{1,0\}$ prediction sets. An option that might be acceptable to all would be to withhold a decision until additional information was obtained that could improve forecasted outcome differentiation. In any case, we have demonstrated with real data on 300,000 arrested offenders at their arraignments that the provisional premise of White advantage compared to Blacks can be exploited to produce a very close approximation of prediction parity.  

Finally, there is no assurance that the comparability shown in Table~\ref{tab:t2} will be achieved in other settings. The number of observations matters. So do the properties of the data. We have produced prediction parity but have not guaranteed it for new data. It remains to be seen how widely our results might generalize. The challenge comes not just from different mixes of offenders, but from different ways to define ``similarly situated.'' Should juvenile arrests, for example, not be used?

\section{Discussion}

The statistical principles used in our fairness formulation are easily generalized. We have already noted that a wide range of classifiers can be used in training and that one is not limited to binary outcomes. Across the range of statistical tools employed, there are also tuning parameters that can be varied depending on the nature of the data and the setting. For example, the value of the coverage probability for conformal prediction sets can trade-off uncertainty against the number of outcome categories likely to be included. 

One is not limited to one disadvantaged, protected class. The analysis will be somewhat more involved, but for instance, one might treat Black offenders and Hispanic offenders as if they were White offenders. The limiting consideration will likely be a subject-matter rationale. Just as for Black offenders, there is substantial heterogeneity among Hispanic offenders (e.g., those with Cuban versus Puerto Rican descent). What matters is what impact that heterogeneity has \emph{on treatment by the criminal justice system and whether there are legal consequences for protected classes.} For example, country of origin might well matter, but probably not marital status. One must not forget that the relevant perceived differences between offenders usually are social constructions. 

One also is not limited to training on a single protected class as the baseline. One can include two or more demographic classes insofar as that can be justified in subject-matter terms. The advantaged demographic class might include White offenders and Asian offenders. One can also be more granular. For example, the advantaged demographic class might be Whites and Asians from affluent zip codes. However, there can be jurisprudential and political issues if the inclusion criteria are novel or contradict accepted practice.

As a technical matter, one can choose any subset of offenders to set the training baseline. However, important features of our approach can be compromised. For example, the outcome on which the training is done should be accepted as reasonably fair or the classifier will build in unacceptable unfairness. Such concerns are relevant even if only some of the training cases are treated unfairly by the criminal justice system. That is one reason why one might choose not to train on both Black and White offenders. 

But the issues can be subtle and difficult to resolve. Might one be prepared to argue, for example, that if offenders were selected solely from affluent zip codes, race no longer matters? Could one sensibly train on affluent Whites and Blacks? And if so, is the range of offenses and covariate values adequate to permit sufficiently accurate risk assessments for individuals from far less affluent zip codes? Gang membership, for instance, might only be a risk factor in low income neighborhoods and would be statistically overlooked if training were undertaken only with offenders from moderate to high income neighborhoods. To repeat a recurring theme, the analysis must be constructively shaped by subject-matter and jurisprudential expertise.

Finally, one can easily imagine improving the data, the classifier, and conformal prediction set tuning such that a risk algorithm for White offenders has enhanced performance. That might lead to increased privilege for Whites and ordinarily could enlarge a White/Black fairness gap. But our approach to fairness would confer those very same advantages on Black offenders. Then, both protected classes can be made better off.

\section{Conclusions}

We have focused on the input and output of algorithmic risk assessments. Although any subsequent decisions or actions may be unfair, they are beyond a risk algorithm's reach and are best addressed by reforms tailored to those phenomena. Blaming a risk algorithm is at best a distraction and can divert remediation efforts away from fundamental change. 

We have shown empirically that, at least for our data, prediction parity as a form of group fairness is easily achieved for a sensible risk baseline by (1) training on a more privileged protected class and then (2) transporting the joint predictor probability distribution from a less privileged protected class to the joint predictor probability distribution of a more privileged protected class. Pareto improvement at the level of protected classes can follow. The more privileged class is not made worse off. This also holds for the less privileged class that can be made better off as well. One might then argue that the risk algorithm dice are no longer loaded to favor one protected class over another. This strikes directly at concerns about mass incarceration and its many consequences. 

By achieving prediction parity, we marshal support for the use of privilege as an organizing framework for internal fairness. We readily acknowledge the many unresolved questions about the precise content and meaning of ``privilege,'' especially in criminal justice settings, but whatever its nature, when a risk algorithm treats Black offenders as if they are White offenders, prediction parity can directly follow. 

At the same time, we can make no claims about external fairness: classification parity, forecasting accuracy parity and cost ratio parity. This is surely disappointing, but simply reflects the limitations of historical data when a fair risk algorithm is being sought. In other recent work (Berk et al, 2022), there are interesting ideas for how progress might made by explicit framing of the risk algorithm's development as a causal intervention in the decision-making status quo.   

We also recognize that by intervening on behalf of members of less privileged protected classes, we concurrently introduce a form of differential treatment. This has a long and contentious fairness history. But in most such circumstances, some classes arguably are made better off as other classes arguably are made worse off. Our approach to criminal justice risk assessment can sidestep such concerns insofar as there is differential advantage that can be exploited. Still, Pareto improvement for protected classes must pass political and legal muster. One hurdle is whether under our approach real and consequential injuries can be avoided (Lujan v. Defenders of Wildlife, 1992); in U.S. federal court, ``injury in fact'' is mandatory. Another hurdle is whether there would be violations of ``equal protection'' under the fifth and fourteenth amendments to the U.S. Constitution (Coglianese and Lehr, 2017: 1191 - 1205), despite our intent make equal protection more equal. We are explicitly using race in our adjustments for fairness by training only on White offenders and transporting the Black covariate distribution to the White covariate distribution. The jurisprudential issues are surprisingly subtle, but the use of race in this manner actually might survive judicial scrutiny (Hellman, 2020).

Finally, there can be concerns about proposing risk procedures, even if rigorous, that explicitly respond to criminal justice realpolitik. But current reform efforts are too often mired in misinformation and factional maneuvering, neither of which improve public discourse. In contrast, our foundational premises are plain. Past research is consulted. The limits of our methods are explicit. And, we have shown with real data that they can be successfully applied. 

\pagebreak
\appendix
\section{Our Fair Risk Algorithm as Pseudocode}\label{appsec:pseudocode}
Algorithm~\ref{alg:fair-risk-alg} presents our fair risk algorithm as a pseudocode.
\begin{algorithm}[!h]
    \SetAlgoLined
    \SetEndCharOfAlgoLine{}
   \KwIn{Data $\mathcal{D} = \mathcal{D}_w \cup \mathcal{D}_b$ where $\mathcal{D}_w$ is the white population data and $\mathcal{D}_b$ is the black population data; Coverage probability $1 - \alpha$.}
    \KwOut{A ``fair'' point prediction and prediction interval for the response.}
    Split the white population data $\mathcal{D}_w$ into two parts: white training data $\mathcal{D}_{w,t}$ and white calibration data $\mathcal{D}_{w,c}$. \;
    Use $\mathcal{D}_{w,t}$, fit a classifier for the response given the covariates. Call this $\widehat{f}^{\texttt{white}}(\cdot)$ that takes an $x$ from the white joint probability distribution and outputs a vector of probabilities $\widehat{f}^{\texttt{white}}(x)$.\;
    Use $\mathcal{D}_{w,c}$ to obtain a conformal prediction set $\widehat{C}_{\alpha}^{\texttt{white}}(\cdot)$ that takes an $x$ from the white joint probability distribution and outputs a set of outcomes that is guaranteed to contain the corresponding white outcome with a probability of at least $1 - \alpha$. See Appendix~\ref{appsec:conformal-inference} and Algorithm~\ref{alg:nested-conformal-prediction} for more details on constructing conformal prediction sets.\;
    Using the covariate observations in $\mathcal{D}_b$ and $\mathcal{D}_{w,t}$, obtain an estimate $\widehat{T}(\cdot)$ of the optimal transport map, that takes a covariate vector $x$ from a black joint probability distribution and outputs $\widehat{T}(x)$ which resembles a white joint probability distribution. See Appendix~\ref{appsec:optimal-transport} and Algorithm~\ref{alg:optimal-transport-map} for more details on estimating the optimal transport map.\;
    \Return the point prediction output of our fair risk algorithm as follows. If $x$ is the covariate vector of a white person, set $\widehat{y}^{\texttt{white}}$ as the highest probability outcome among $\widehat{f}^{\texttt{white}}(x)$. If $x$ is the covariate vector of a black person, set $\widehat{y}^{\texttt{black}}$ as the highest probability outcome among $\widehat{f}^{\texttt{white}}(\widehat{T}(x))$. \;
    \Return the prediction set output of our fair risk algorithm as follows. If $x$ is a covariate vector of a white person, return $\widehat{C}_{\alpha}^{\texttt{white}}(x)$. If $x$ is a covariate vector of a black person, return $\widehat{C}_{\alpha}^{\texttt{white}}(\widehat{T}(x))$. 
   \caption{``Fair'' Risk Algorithm}
    \label{alg:fair-risk-alg}
\end{algorithm}
 \section{An Introduction to Optimal Transport}\label{appsec:optimal-transport}
Suppose we have two distributions $P$ and $Q$. In our example, think of $P$ as the distribution of covariates for a population Black offenders and $Q$ as that distribution for a population of White offenders. We seek to ``transport'' $P$ to $Q$. Given a random vector $X$ from the distribution $P$, we wish to create $Y = T(X)$ such that $Y$ has the distribution $Q$. The function $T(\cdot)$ is called a transport map taking $P$ to $Q$. There commonly exists several such maps, but (under regularity conditions) there is a unique map that minimizes the distance between $X$ and $T(X)$ while ensuring $T(X)$ has the $Q$ distribution. In other words, $T(X)$ moves $X$ as little as possible to approximate a random vector from $Q$. Such a unique map, denoted by $T^*(\cdot)$, is called the optimal transport map. 

Several kinds of distances can be used. Probably, the most common is the Euclidean distance and with this choice, the optimization problem known at the Monge formulation, defines the optimal transport map $T^*(\cdot)$ given by
\[
T^* ~:=~ \argmin_{\substack{T:\,T(X)\sim Q,\\\mbox{if }X\sim P}}\,\mathbb{E}_P[\|X - T(X)\|_2^2].
\]
This means that $T^*$ is the minimizer of $\mathbb{E}_P[\|X - T(X)\|_2^2]$ over all functions $T$ such that $T(X) \sim Q$ whenever $X\sim P$. This constraint on the functions $T$ ensures that $T^*$ transports $P$ to $Q$ and minimizes the expectation, which ensures that it is an optimal in that sense. 

To illustrate, we use two very simple, univariate distributions: $P$ and $Q$ each supported on $5$ points. Distribution $P$ is supported at $6, 10, 15, 20, 25$, and distribution $Q$ is supported at $10, 12, 15, 20, 30$. The probability values are given by
\begin{align*}
P(6) = P(10) = P(15) = P(20) = P(25) &= {1}/{5},\\
Q(10) = Q(12) = Q(15) = Q(20) = Q(30) &= {1}/{5}.
\end{align*}
Consider two transport maps $T_1$ and $T_2$ that convey values from $\{6, 10, 15, 20, 25\}$ into $\{10, 12, 15, 20, 30\}.$
\begin{align*}
T_1(6) = 10,\, T_1(10) = 12,\, T_1(15) = 15,\, T_1(20) = 20,\, T_1(25) = 30,\\
T_2(6) = 12,\, T_2(10) = 10,\, T_2(15) = 15,\, T_2(20) = 20,\,T_2(25) = 30.
\end{align*}
In words, $T_1$ matches the smallest in the support of $P$ to the smallest in the support of $Q$, the second smallest in the support of $P$ to the second smallest in the support of $Q$, and so on. In this example, the transport map is also the quantile-quantile map. On the other hand, $T_2$ does not change the same values (i.e., 10 to 10, 15 to 15, 20 to 20). It is easy to verify that if $X$ has the distribution $P$, then $T_1(X)$ and $T_2(X)$ both have the distribution $Q$. This illustrates that a map transporting two general distributions $P$ to $Q$ is not unique. In this example, $T_1$ is the optimal transport plan minimizing $\mathbb{E}_{P}[|X - T(X)|^2]$, where the expectation is with respect to $X$ from the distribution $P$, over all maps $T$ such that $T(X)$ has the distribution $Q$. 

For any transport map $T$, writing $Y = T(X)$, we obtain a joint distribution for the augmented vector $(X, Y) = (X, T(X))$, which is a ``coupling'' between the distributions $P$ and $Q$. From this coupling perspective, the problem of optimal transport can be reformulated in terms of finding that coupling for a joint distribution whose marginals are fixed at $P$ and $Q$. This is called the Kantorovich formulation. Estimation of the optimal transport plan is undertaken given data from $P$ and $Q$, \emph{not} the distributions $P$ and $Q$ themselves. We will not provide more details, and refer the reader to~(Peyr\'e and Cuturi, 2019; Deb and Sen, 2021; Deb et al., 2021; H\"utter and Rigollet, 2021).

We offer pseudocode below (Algorithm~\ref{alg:optimal-transport-map}) for estimating the optimal transport map $T^*$ based on data, drawing on Sections 6.1.1 and 6.1.3 of H\"utter and Rigollet (2021) with minor differences.
\begin{algorithm}[htp]
    \caption{Estimation of optimal transport map}
    \label{alg:optimal-transport-map}
    \SetAlgoLined
    \SetEndCharOfAlgoLine{}
    \KwIn{Data $\mathcal{D}_1 = \{X_1, \ldots, X_m\}$ from distribution $P$ (in dimension $d$) and data $\mathcal{D}_2 = \{Y_1, \ldots, Y_n\}$ from distribution $Q$ (in dimension $d$).}
    \KwOut{A transport map $\widehat{T}(\cdot)$.}
    Find 
    \begin{equation}\label{eq:definition-Gamma}
    \widehat{\Gamma} := \argmin_{\substack{\Gamma\in\mathbb{R}^{m\times n}, \Gamma_{ij} \ge 0\\\sum_{i=1}^m \Gamma_{ij} = 1/n, 1\le j\le n,\\ \sum_{j=1}^n \Gamma_{ij} = 1/m, 1\le i\le m}}\,\sum_{i=1}^m \sum_{j=1}^n \|X_i - Y_j\|_2^2\Gamma_{ij}.
    \end{equation}
    This is a linear programming problem and can be solved using the R package \texttt{transport}.\;
    
    For each $1\le i\le m$, define
    \[
    \widehat{Y}_i := \widehat{T}^{\mathrm{emp}}(X_i) = \frac{\sum_{j=1}^n \widehat{\Gamma}_{ij}Y_j}{\sum_{j=1}^n \widehat{\Gamma}_{ij}} = m\sum_{j=1}^n \widehat{\Gamma}_{ij}Y_j,
    \]
    as the transport of $X_i$ (observations in $\mathcal{D}_1$).\;
    
    For $1\le k\le d$, perform non-parametric regression (using kernels, random forest, RKHS, etc) on the data $(X_i, \widehat{Y}_{i,k}), 1\le i\le m$ with the $k$-th coordinate of $\widehat{Y}_i$ as the response. This yields a map $\widehat{T}_k(\cdot)$.\;
    \Return $\widehat{T}(x) := (\widehat{T}_1(x), \ldots, \widehat{T}_d(x))\in\mathbb{R}^d$ for any $x\in\mathbb{R}^d$ as the transport of $x$. This map $\widehat{T}(\cdot)$ serves as an estimate of the optimal transport map that transports $P$ to $Q$.
\end{algorithm}
The problem~\eqref{eq:definition-Gamma} is a linear programming task and is the most computing-intensive part of Algorithm~\ref{alg:optimal-transport-map}. Beyond several thousand observations, solving~\eqref{eq:definition-Gamma} is computationally prohibitive. A simple work around is to split the data into several parts, apply Algorithm~\ref{alg:optimal-transport-map} on each part, and then average the estimates of optimal transport thus obtained. 

Formally, suppose $\mathcal{D}_1^*$ and $\mathcal{D}_2^*$ are the initial (big) datasets available from $P$ and $Q$ respectively. Split $\mathcal{D}_1^*$ randomly into, say, 10 batches. Call them $\mathcal{D}_{1,1}, \mathcal{D}_{1,2}, \ldots, \mathcal{D}_{1,10}$. Similarly, split $\mathcal{D}_2^*$ randomly into, say, 10 batches. Call them $\mathcal{D}_{2,1}, \mathcal{D}_{2,2}, \ldots, \mathcal{D}_{2,10}$. Apply Algorithm~\ref{alg:optimal-transport-map} on $\mathcal{D}_{1,1}, \mathcal{D}_{2,1}$ to obtain an estimate $\widehat{T}^{(1)}(\cdot)$ of the optimal transport map. Similarly, apply Algorithm~\ref{alg:optimal-transport-map} on $\mathcal{D}_{1,2}, \mathcal{D}_{2,2}$ to obtain $\widehat{T}^{(2)}(\cdot)$, and so on to obtain $\widehat{T}^{(3)}(\cdot), \ldots, \widehat{T}^{(10)}(\cdot)$. Because the datasets $\mathcal{D}_{1,j}, \mathcal{D}_{2,j}$ are of sizes 10 times smaller than $\mathcal{D}_1*, \mathcal{D}_2^*$, problem~\eqref{eq:definition-Gamma} becomes more manageable computationally. Finally, set for all $x\in\mathbb{R}^d$,
\[
\widehat{T}(x) := \frac{1}{10}\sum_{j=1}^{10} \widehat{T}^{(j)}(x),
\]
as an estimate of the optimal transport map. For concreteness, here the data is split into 10 batches, but it can be made into a larger number of batches as long as each batch contains ``enough'' observations.  

\section{An Introduction to Conformal Inference}\label{appsec:conformal-inference}
In the context of regression or classification, prediction can be a major goal, and there exist several point prediction methods that report an estimate of the true response. In practice, it is often important to also provide an uncertainty quantification along with the point prediction. Conformal inference provides such uncertainty quantification. In the context of our data, we only provide details about conformal inference for classification. 

The setting of classical conformal inference is as follows. One has observations $(X_1, Y_1), \ldots, (X_n, Y_n)$ independent and identically distributed from a distribution $P$. The goal is to provide a set $\widehat{C}_{\alpha}$ such that
\begin{equation}\label{eq:prediction-coverage}
\mathbb{P}((X_{n+1}, Y_{n+1}) \in \widehat{C}_{\alpha}) \ge 1 - \alpha,
\end{equation}
when $(X_{n+1}, Y_{n+1})\sim P$. Conformal inference provides a set $\widehat{C}_{\alpha}$ that satisfies~\eqref{eq:prediction-coverage} without any assumptions on the underlying joint probability distribution $P$. Furthermore, use can be made of one's favorite prediction algorithm, and the validity guarantee holds regardless of what the algorithm employed. 

The basic idea of conformal inference is to assign a real valued score to each of the calibration data points, and a future point is placed in the prediction set if its score ``conforms'' with those of the training data points. There are several ways to construct such scores and all of them lead to a valid prediction set. In the following, we describe a simple score, and more complicated score, such as those in Kuchibhotla and Berk (2021), can also be used to obtain more precise prediction sets. 

For a conformal inference procedure for classification, consider a setting in which the response/outcome $Y_i$ takes one of two values $0$, and  $1$. The pseudocode for the conformal inference with an absolute residual score is given in Algorithm~\ref{alg:nested-conformal-prediction}. In the context of our fair risk algorithm
(as shown in Figure~\ref{fig:flowchart2}), Algorithm~\ref{alg:nested-conformal-prediction} can be applied from Step 2, because the classifier $\widehat{p}(\cdot|\cdot)$ is given by $\widehat{f}^{\mathrm{white}}(\cdot)$. In other words, $\mathcal{D}_{w,t}$ and $\mathcal{D}_{w,c}$ in Figure~\ref{fig:flowchart2} play the roles of $D_1$ and $D_2$, respectively, in Algorithm~\ref{alg:nested-conformal-prediction}. Further, $\widehat{f}^{\mathrm{white}}(x)$ plays the role of $(\widehat{p}(0|x), \widehat{p}(1|x))$ in Algorithm~\ref{alg:nested-conformal-prediction}.

\begin{algorithm}[h]
    \caption{Conformal prediction for classification}
    \label{alg:nested-conformal-prediction}
    \SetAlgoLined
    \SetEndCharOfAlgoLine{}
    \KwIn{Data splits $D_1$ (training data) and $D_2$ (calibration data), coverage probability $1 - \alpha$.}
    \KwOut{A prediction set $\widehat{C}_{\alpha}(\cdot)$ such that $\mathbb{P}(Y_{\mathrm{f}}\in\widehat{C}(X_{\mathrm{f}})) \ge 1 - \alpha$, for a future observation $(X_{\mathrm{f}}, Y_{\mathrm{f}})$.}
    Train a classifier $\widehat{p}(\cdot|\cdot)$ on the training data $D_1$. This gives a probability distribution (estimator) for the outcomes for each $x$, i.e., we get for each $x$, probabilities $\widehat{p}(0|x)$ and $\widehat{p}(1|x)$ such that $\widehat{p}(0|x) + \widehat{p}(1|x) = 1$.\;
       For each $(X_i, Y_i)$ in the calibration data $D_2$, calculate the conformal scores $s(X_i, Y_i)$ as follows: 

\[
s(X_i, Y_i) := |Y_i - \widehat{p}(Y_i|X_i)|.
\]
\;
    \vspace{-0.1in} 
    Compute the $(1 + 1/|D_2|)(1 - \alpha)$-th quantile of $s(X_i, Y_i), i\in D_2$. Call this quantile $\widehat{\gamma}(\alpha)$.\;
    \Return the prediction set 
    \begin{equation}\label{eq:nested-predict-set-algo}
    \widehat{C}_{\alpha}(x) ~:=~ \left\{y\in\{0, 1\}:\,s(x, y) = |y - \widehat{p}(y|x)|\le\widehat{\gamma}(\alpha)\right\}.
    \end{equation}
\end{algorithm}

\section{Conditions Under which Internal Fairness Implies External Fairness}\label{appsec:external fairness}
Within our fairness formulation, for a risk algorithm used at arraignments to demonstrate external fairness, an appropriate estimate of the probability of a post-arraignment arrest must be available. Estimates can be obtained from the test data on hand, but they characterize criminal justice business as usual. Such estimates are appropriate for White offenders but not for the counterfactual of Black offenders who, post-release, are treated by police the same as similarly situated White offenders.\footnote
{
We focus on police because in practice, it is police who decide whether to make an arrest. Arrests by citizens are permitted but are extremely rare. Also, we use the term ``offender'' throughout to be consistent with our arraignment application.
}  
In this appendix, we provide sufficient conditions, under the counterfactual, allowing external fairness for classification parity, forecasting accuracy parity, and cost ratio parity to be properly inferred from internal fairness estimates. To this end, we introduce notation needed to address counterfactual outcomes. 

Just in our application, there are two possible post-arraignment outcomes for offenders who are not detained: an arrest for a crime of violence or no such arrest.
Let $ R \in \{w,b\}$ denote the race of an offender, say White ($w$) or Black ($b$), and define $Y(r)$ as the counterfactual outcome a person would experience if, contrary to fact, the person were treated by police as a person of race $r=\{b,w\}$. Consistent with current causal inference thinking, each person, irrespective of his or her race, is associated with a pair of counterfactual outcomes ${Y(w),Y(b)}$, depending on how they would be treated by police were they of a given race, including a race which hypothetically differs from their actual race. That is, a person of race $b$ is treated as a person for race $w$ or vice versa. 

Incorporating covariates, let $Y(r,x)$ likewise denote the counterfactual outcome had the person been treated by police as a person of race $r=b,w$ with covariate values $x$. In principle, covariates $X$ may be multivariate and may include both continuous or discrete variables. For instance an offender's age usually is a key covariate to consider. Each offender in the data has a reported age: $X=$ Age. Suppose for a 20 year old Black offender there are well-defined counterfactual outcomes $\{Y(w,x):x=21,22,\ldots,50\}$ corresponding to the person's outcomes if, contrary to fact, police treated this individual as White and of age 21, or 22, \ldots, or 50 years old. We will see shortly that the relationship between counterfactual variables $Y(r)$ and $Y(r,x)$ can be subtle.  

Furthermore, for each Black offender, we let $Y(r,X(r^*))$ denote the person's counterfactual outcome had the offender been treated by the police as if he were of race $r$, with covariate values set to what they would have been had the offender been of race $r^*$.  For example, suppose $X$ includes age and number of prior arrests, and set $r=r^*=w$. Then, the corresponding counterfactual $Y(r,X(r^*))=Y(w,X(w))$ for a Black offender 26 years of age with 2 prior burglary arrests such that $X=(26,2)$ defines his or her outcome were the Black offender treated by police as if the offender were White with an age and number of prior burglaries corresponded to a similarly situated white person, such as $X(w)=(22, 1)$. For the Black offender, therefore, there is a counterfactual $Y(w,X(w))=Y(w,X(w)=(22,1))$. This follows from our intent to treat Black offenders as if they were White.

Throughout, we make the following consistency assumptions, which provide a necessary link between various defined counterfactuals. Mainly, we assume that $Y=Y(r)=Y(r,X(r))$ almost surely, if $R=r$. The observed outcome in the test data for an offender of race $R=r$, matches the hypothetical outcome the offender would have had were the police to treat the offender as a person of race $R=r$, which in turn matches the offender's potential outcome if the police were to treat him or her as a person of race $R=r$ and covariates $X(r)=x$. Consistency may fail to hold, for instance, if a black offender's outcome depends on another offender's race or covariate values in addition to his own. This may arise in settings where an offender's arrest may not only depend on the offender's race but also on the race of an accomplice. Then, the potential outcome for the black offender may be ill-defined unless the race of the accomplice is introduced as a covariate.     
Formally, suppose our proposed algorithm aspires to forecast the counterfactual $Y(w,X(w))$ for a Black offender (i.e., conditional on $R=b$) with observed covariates $X=X(b)$. Consider the following condition linking the optimal transport map $T^*$ to counterfactual outcomes:
\begin{equation}\label{eq:counterfactual Optimal Transport}
X(w)=T^*(X(b)) \hspace{.1in} w.p.1,
\end{equation}

The assumption essentially states that for each Black offender, the joint distribution of $X(w)$ and $X(b)$ is degenerate. The assumption is far from trivial, as illustrated in the simple case where the optimal transport map is a location shift, i.e., $T^*(x) = x - \mu$ for a fixed constant $\mu$. A violation of the assumption can arise in this setting if there were a covariate Z, say whether the offender had a family member who had been incarcerated, that although not observed in the database, interacts with race to modify the person's potential outcome as follows: $X(w)=T^*(X(b))=X(b)-\mu_0-\mu_1 \times Z$. Failing to account for $Z$ would invalidate the equality assumption because the relationship between the two potential outcomes $X(w)$ and $X(b)$ cannot be made deterministic unless one also conditions on the unobserved factor $Z$. 

Finally, consider the following strong ignorability condition, that for $r,r^* \in {w,b}$ \begin{equation}\label{eq:Joint Ignorability}
Y(r,x) \perp \!\!\! \perp R,X(r^*),
\end{equation}
which states that there are no common factors that determine both whether a person of race $R$ and covariates $X(r^*)$ had he been of race $r^*$ interacts with the criminal justice system, and how that system would treat offeneder if he or she were of race $r$ with covariates $x$.

Under conditions~\eqref{eq:counterfactual Optimal Transport} and~\eqref{eq:Joint Ignorability}, we prove that
\[
Y(w, T^*(X(b)))\big|R = b, X \;\overset{d}{=}\; Y(w, X(w))\big|R = w, X.
\]

This follows by noting that 
for any $x$ and $x^*=T^*(x)$:
\begin{align*}
&\mathbb{P}(Y(w,x^*)=y,X(w)= x^*|R=b, X=x)\\
&=\mathbb{P}(Y(w,x^*)= y|R=b, X(b)=x)Pr(X(w)=x^*|R=b,X(b)=x)\\
&=\mathbb{P}(Y(w,x^*)= y|R=b, T^*X(b)=x^*)) \times 1\\
&=\mathbb{P}(Y(w,x^*)= y|R=b, X(w)=x^*))\\
&=\mathbb{P}(Y=y|X=x^*,R=w) 
\end{align*}
establishing the result that the desired distribution $(Y^*=Y(w,x^*),X(w))|R=b$ can be obtained by first sampling $T^*(X)$ from the covariate distribution of black defenders, and subsequently sampling $Y$ from the conditional distribution of white defenders with covariate equal to $T^*(X)$, matching the output of our proposed algorithm. Should this hold, external fairness can be said to be achieved if internal fairness can be established for the proposed algorithm. More precisely, to the extent that classification, forecasting accuracy and cost ratio parities can be demonstrated, their counterfactual analogues would in principle be implied by the stated assumptions. One would then have a firm basis for claiming external fairness. 

There are several ways in which these assumptions arguably are unrealistic, at least in most jurisdictions in the United States.  First, as explained above, the assumption that the relationship between counterfactual is deterministic rules out the existence of latent effect heterogeneity in the association between an offender's race and the manner in which the offender is ultimately treated by the criminal justice system. This assumption is sometimes called a rank preservation condition, which  implies that for any two persons $i$ and $j$, if $X_i(b) \leq X_j(b)$ for a scalar variable $X$, then it must be that $X_i(w) \leq X_j(w)$, thus ruling out the existence of an unmeasured factor related to race in a manner that can alter the ranking of potential covariate values. 
For the location shift example previously described with say $\mu_0=0$ and $\mu_1=5$, then $X(w)=T^*(X(b))=X(b)-5 \times Z $. It is possible that $X_i(b)=4 \leq X_j(b)=6$. However $Z_i=0$ while $Z_j=1$ so that $X_j(w)=1 \leq X_i(w)=4$. The rank preservation is violated. Even in this simple example, rank preservation assumption, which is not empirically testable, may be difficult to justify because of a large number of omitted variables, particularly in practical settings where $X$ is multivariate.

The independence condition~\eqref{eq:Joint Ignorability} is likewise unrealistic because it rules out any common factor associated with the race of an offender and a potential apprehension outcome. It also rules out any unmeasured common cause of an offender's covariates and an apprehension outcome, conditional on race. Given the role of race in myriad social institutions and interactions,
claims that such relationships are absent would in practice strain credibility.   

\section{Confusion Table Results for Black Test Data When the Response Variable as well as The Predictors are Transported}

Stakeholders and others often focus primarily on fairness represented in confusion tables from test data. In deference to those individuals, we applied optimal transport to the Black test data in a manner that included in the joint probability distribution the response variable as well as the predictors. Table~\ref{tab:t2} is virtually the same as Table~\ref{tab:t1} within sampling error. The comparability is striking. For example, we compared the base rates from the test data for White offenders and the transported test data for Black offenders. For both, the base rate was approximately .075. 7.5\% of both Black and White offenders were rearrested after an arraignment for a violent crime. We emphasize the base rate equivalence because the base rate is so central in formal fairness discussions (Kleinberg et al., 2017).

\begin{table}[htp]
\scriptsize
\caption{Transported Test Data Confusion Table for Black Offenders Using White-Trained Algorithm (30\% Predicted to Fail, 7.5\% Actually Fail)}
\begin{center}
\begin{tabular}{|c|c|c|c|}
\hline 
 Actual Outcome & No Violence Predicted &Violence Predicted & Classification Error  \\ 
 \hline 
 No Violence & 2658 & 1042 (false positive) & .28 \\
 Violence & 135 (false negative) & 166 &  .45 \\
 \hline
 Forecasting Error & .05 & .86 &  \\
 \hline \hline
\end{tabular}
\end{center}
\label{tab:t2}
\end{table}

If one is prepared, as in common practice, to evaluate fairness solely using the test data in a confusion table, optimal transport provides an effective equalizer. One may have politically acceptable risk algorithm. One is also externally fair for predictive parity because no outcome label is required. And, if one is comfortable assuming that in reality, Black offenders will be treated on the average the same as similarly situated White offenders after an arraignment release, external fairness is achieved more generally. But for most stakeholders, this last step will stretch credibility. 

In short, although interpretations for fairness will vary, one has made all of the confusion table results for White offenders and Black offenders the same. All of the trade-off concerns are bypassed as long as one is prepared to assume that if a confusion table is good enough for White offenders, it is good enough for Black offenders. Note that such claims go only to the aggregate confusion table results by which one might evaluate a risk algorithm overall. It says nothing about the operational issues required for forecasting. 

\pagebreak
\section*{References}
\begin{description}
\item
Allen, M., and Crook, J. (2017) ``More Than Race: Proliferation of Protected Groups and the Increasing Influence of the Act.'' in G.D. Squires (ed.) \textit{The Fight for Fair Housing} Routledge Press. \item
Alpert, G.P., Dunham, R.G., and M.R. Smith (2007) ``Investigating Racial Profiling by the Maimi-Dade Police Department: A Multimethod Approach.'' \textit{Criminology and Public Policy} 6(1) 24 -- 55.
\item
Baer, B.R., Gilbert, D.E., and M.T.Wells (2020) `` Fairness Criteria through the Lens of Directed Acyclic Graphs: A Statistical Modeling Perspective.'' In Dubber, M.D., Pasquale, F., and S.Das, \emph{The Oxford Handbook of Ethics of AI}. Oxford Press.
\item
Barocas, S., Hardt, M.,  and A. Narayanan (2018) \textit{Fairness and Machine Learning}. http://www.fairmlbook.org
\item 
Bekbolatkyzy, D.S., Yerenatovna, Yergali, D.R., Maratuly, Y.A., Makhatovna, A.G., and K.M. Beaver. (2019) ``Aging Out of Adolescent Delinquency: Results from a Longitudinal Sample of Youth and Young Adults.'' \emph{Journal of Criminal Justice} 60 (January - February): 108 -- 116.
\item
Berk, R.A. (2009) ``The Role of Race in Forecasts of Violent Crime,'' \textit{Race and Social Problems}, 1(4): 231--242.
\item
Berk, R.A. (2017) ``An Impact Assessment of Machine Learning Risk Forecasts on Parole Board Decisions and Recidivism.'' \textit{Journal of Experimental Criminology} 13: 193--216.
\item
Berk, R.A. (2018) \textit{Machine Learning Forecasts of Risk in Criminal Justice Settings}. New York: Springer.
\item
Berk, R.A., Heirdari, H., Jabbari, S., Kearns, M., and A. Roth (2018) ``Fairness in Criminal Justice Risk Assessments: The State of the Art.'' \textit{Sociological Methods and Research}, first published July 2nd, 2018, http: //journals.sagepub.com/doi/10.1177/0049124118782533.
\item
Berk, R.A., and A. A. Elzarka (2020) ``Almost Politically Acceptable Criminal Justice Risk Assessment.'' \textit{Criminology and Public Policy} 2020: 1 -- 28.
\item
Berk, R.A., Kuchibhotla, A.K., and Tchetgen Tchetgen, E. (2022)
``Fair Risk Algorithms.'' \textit{Annual Review of Statistics and Its Applications}, in press.
\item
Bhopal, K. (2018) \textit{White Privilege: The Myth of a Post-Racial Society}. Policy Press.
\item
Binns, R. (2020) ``On the Apparent Conflict between Individual and Group Fairness,'' FAT* '20: Proceedings of the 2020 Conference on Fairness, Accountability, and Transparency, January 2020 Pages 514 -- 524.
\item
Breiman, L. (2001) ``Statistical Modeling: Two Cultures (with comments and a rejoinder by the author,'' \textit{Statistical Science} 16(3) 199 -- 231.
\item
Chouldechova, A. (2017) ``Fair Prediction with Disparate Impact: A Study of Bias in Recidivism Prediction Instruments.'' \textit{Big Data} 5(2):153 -- 163.
\item
Coglianese, C., D. Lehr (2017) ``Regulating by Robot: Administrative Decision Making in the Machine-Learning Era.'' \textit{Georgetown Law Journal} 105: 1147--
\item
Coglianese, C., D. Lehr (2019) ``Transparency and Algorithmic Governance.'' \textit{Faculty Scholarship at Penn Law} 2123. Also \textit{Administrative Law Review} 2019 1 (2019).
\item
Coston, A., Ramamurthy, K.N., Wei, D., Varshney, K.R., Speakman, s., Mustahsan, A., and Chakraborty, S. (2019) ``Fair Transfer Learning with Missing Protected Attributes.'' \textit{AIES}'19: 91-98, Section 1, Algorithmic Fairness.
\item
Corett-Davies, S., Pierson, E., Feller, A., Goel, S., and A. Huq (2017) ``Algorithmic Decision making and the Cost of Fairness,'' \textit{KDD '17: Proceedings of the 23rd ACM SIGKDD International Conference on Knowledge Discovery and Data Mining}: 797 -- 806.
\item
Corbett-Davies S., and S. Goel (2018) ``The Measure and Mismeasure of Fairness:  A Critical Review of Fair Machine Learning.'' 35th International Conference on Machine Learning (ICML 2018). 
\item
D'Amour, A., Heller, K., Adlam, B., et al., (2020) ``Underspecification Presents Challenges for Credibility in Modern Machine Learning.'' arXiv:2011.03395v2 [cs.LG].
\item
De Lara, L., Gonz\'alez-Sanz, A., Asher, N., and J.-M. Loubes (2021) ``Counterfactual Models: The Mass Transportation Viewpoint'' Institut de Recherche en Informatique de Toulouse. https://www.irit.fr/
\item
Diana, E., Gill, W., Kearns, M., Kenthapadi, K., and A. Roth (2021) Minimax Group Fairness: Algorithms and Experiments) AIES' 21: Proceedings of the 2021 AAI/ACM: 66--76.
\item
Dwork, C., Hardt, M., Patassi, T., Reingold, O., and R. Zemel (2012) ``Fairness through Awareness.'' ITCS 2012: Proceedings of the 3rd Innovations in Theoretical Computer Science Conference: 214 -- 226.
\item
Edwards, F., Lee, H., and M. Esposito (2019) ``Risk of Being Killed by Police Use of Force in the United States by Age, Race-Ethnicity, and Sex.'' Proceedings of the National Academy of Sciences: 116: 16793--16798.
\item
Feldman, M., Friedler, S., Moeller, J., Scheidegger, C., and S. Venkatasubrtamanian (2015)  ``Certifying and Removing Disparate Impact.'' In Proceedings of the 21th ACM SIGKDD International Conference on Knowledge Discovery and Data Mining. ACM, 259 -- 268.
\item
Fisher, F.M., and J.B. Kadane (1983) ``Empirically Based Sentencing Guidelines and Ethical Considerations.'' In Alfred Blimstein et. al., ed. \textit{Research on Sentencing: The Search for Reform, Volume II}, National Criminal Justice Reference Service, Office of Justice Programs, U.S. Department of Justice, NCJ-91771.
\item
Freedman, D.A. (2012) \textit{Statistical Models: Theory and Practice}, revised edition. Cambridge University Press.
\item
Friedman, J.H. (2001) ``Greedy Function Approximation: A Gradient Boosting Machine.'' \textit{Annals of Statistics} 29 (5): 1189 -- 1232. 
\item
Gastwirth, J.L. (2000) \textit{Statistical Science in the Courtroom} Springer.
\item
GBD 2019 Police Violence Subnational Collaborators (2021) ``Fatal Police Violence by Race and State in the USA, 1980-2019: A Network Meta-Regression.'' \textit{Lancet} 398: 1239-1255. 
\item
Gelman, A., Fagan, J., and A. Kiss (2012) ``An Analysis of the New York City Police Department's `Stop-and-Frisk' Policy in the Context of Claims of Racial Bias.'' \textit{Journal of the American Statistical Association} 102 (2007): 813 -- 823
\item
Grogger, J., and G. Ridgeway (2012) ``Testing for Racial Profiling in Traffic Step From Behind a Veil of Darkness.'' \textit{Journal of the American Statistical Association} 202 (2006): 878 -- 887.
\item
''Gupta, C., Podkopaev, A., and Ramdas, A. (2020)``Distribution-Free Binary Classification: prediction sets, confidence intervals, and Calibration.'' arXiv:2006.10564 [stat.ML].
\item
Harcourt, B.W. (2007) \textit{Against Prediction: Profiling, Policing, and Punishing in an Actuarial Age}. Chicago, University of Chicago Press.
\item
Hardt, M., Price, E., N. Srebro (2016) ``Equality of Opportunity in Supervised Learning.'' In D.D. Lee, Sugiyama, U.V. Luxburg, I.  Guyon, and R. Garnett (eds.) \textit{Equality of Opportunity in Supervised Learning}. Advances in Neural Information Processing Systems 29: Annual Conference  on Neural Information Processing Systems 2016, December 5-10, 2016, Barcelona, Spain, (pp.3315 -- 3323).
\item
Hellman, D. (2020) ``Measuring Algorithmic Fairness.'' \textit{Virginia Law Review} 106(4): 811 -- 866.
\item
Holewinski, I.A. (2002) ``Inherently Arbitrary and Capricious: An Empirical Analysis of Variations Among Death Penalty Statutes.'' \textit{Cornell Journal of Law and Public Policy} 12: 231 -- 259.
\item
Horder, J. (1993) ``Criminal Culpability: The Possibility of a General Theory.'' \textit{Law and Philosophy} 12: 193-- 215.
\item
Hudson, B. (1989) ``Discrimination and Disparity: The Infuence of Race on Sentencing.'' \emph{Journal of Ethnic and Migration Studies} 16(1): 23 -- 34.
\item
H\"utter, J., and P. Rigollet (2020) ``Minimax Estimation of Smooth Optimal Transport Maps.'' \textit{Annals of Statistics} 49(2): 1166 -- 1194.
\item
Huq, A.Z. (2019) ``Racial Equality in Algorithmic Criminal Justice.'' \textit{Duke Law Journal} 68 (6), 1043--1134.
\item
Imai, K., and Z. Jaing (2021) ``Principal Fairness for Human and Algorithmic Decision-Making.'' arXiv:2005.10400v4 {cs.CY}
\item
Jackson, E.K. (2019) \textit{Scandinavians in Chicago: The Origins of White Privilege in Modern America} University of Illinois Press.
\item
Johndrow, J.E., and K. Lum (2019) ``An Algorithm for Removing Sensitive Information: Application to Race-Independent Recidivism Prediction.'' \textit{Annals of Applied Statistics} 13(1): 189 -- 220.
\item
Kamiran, F., and T. Calders (2012) ``Data Preprocessing Techniques for Classification Without Discrimination.'' \textit{Knowledge Information Systems} 33:1 - 33.
\item
Kearns, M and A. Roth (2020) \textit{The Ethical Algorithm} Oxford Press.
\item
Kearns, M., Neel, S., Roth, A., and S. Wu (2018) ``Preventing Fairness Gerrymandering: Auditing and Learning for Subgroup Fairness.''\textit{Proceedings of Machine Learning Research} 80: 2564 -- 2572.
\item
Kim, P. (2022) ``Race-Aware Algorithms: Fairness, Nondiscrimination and Affirmative Action.'' \textit{Calfornia Law Review}, forthcoming.
\item
Kleinberg, J., Mullainathan, S., and M. Raghavan, M. (2017) ``Inherent Tradeoffs in the Fair Determination of Risk Scores.'' Proc. 8th Conference on Innovations in Theoretical Computer Science (ITCS).
\item
Kroll, J.A., Huey, J., Barocas, S., Felten, E.W., Reidenberg, J.R., Robinson, D.G., and H. Yu (2017) ``Accountable Algorithms.'' \textit{University of Pennsylvania Law Review} 165 (3): 633 -- 705.
\item
Kuchibhotla, A.K. and R.A. Berk (2021) ``Nested Conformal Prediction Sets for Classification with Applications to Probation Data.''  arXiv:2104.09358
\item
Kushner, H.J., and G.G. Yin (2003) \textit{Stochastic Approximation and Recursive Algorithms and Appications}. Springer.
\item
Kusner, M., Loftus, J., Russell, C. and R. Silva (2017) ``Counterfactual Fairness.'' \textit{Advances in Neural Information Processing Systems 30} (NIPS 2017): 4066 -- 4076 
\item
Lee, D., Huang, X., Hassani, H., and Dobriban, E. (2022) ``T-Cal: An Optimal Test for the Calibration of Predictive Models.'' arXiv:2203.01850v2 [stat.ML]
\item
Lee, N.T., Resnick, P., and G. Barton (2019) ``Algorithmic Bias Detection and Mitigation: Best Practices and Polices to Reduce Consumer Harms.'' Brookings institution, Washington D.C., Bookings Report.
\item
Leonard, D.J., (2017) \textit{Playing While White: Privilege and Power On and Off The Field}. University of Washington Press.
\item
Louffler, C.E., and A. Chalfin (2017) ``Estimate the Crime Effects of Raising the Age of Majority.'' \textit{Criminology \& Public Policy} 16(1): 45 -- 71. 
\item
Luenberger, D.FG. and Y. Ye (2008) \textit{Transportation and Network Flow Problems}. Springer.
\item
Lujan v. Defenders of Wildlife, 504, U.S. 555 (1992).
\item
Lynch, M. (2011) ``Mass Incarceration, Legal Change, and Locale: Understanding and Remediating American Penal Overindulgence.'' \textit{Criminology \& Public Policy} 10(3): 673 -- 698.
\item
Madras, D., Pitassi, T., and R. Zemel (2018) ``Predict Responsibly: Improving Fairness and Accuracy by Learning to Defer.'' \textit{32nd Conference on Neural Information Processing Systems} (NIPS 2018), Montr\'{e}al, Canada.
\item
Madras, D., Creager, E., Pitassi, T., and R. Zemel (2019) ``Fairness through Causal Awareness: Learning Causal Latent-Variable Models for Biased Data.'' \textit{FAT* '19: Proceedings of the Conference on Fairness, Accountability, and Transparency}: 349 --358. 
\item
Manole, T., Balakrishnan, S., Niles-Weed, J., and L. Wasserman (2021) ``Plugin Estimation of Smooth Optimal Transport Maps.'' asXiv:2107.12364v1 [math.ST]
\item
Mason, S.G., (2019) ``Bias in, Bias Out.'' \textit{Yale Law Journal} 128(8): 2219 -- 2298.
\item
Mishler A., and E. Kennedy (2021) ``FADE: Fair and Double Ensemble Learning for Observable and Counterfactual Outcomes.'' arXiv:2109.00173v1. [stat.ML]
\item
Mitchel, S., Potash, E., Barocas, S., D'Amour, A, and K. Lum (2021) ``Algorithmic Fairness: Choices, Assumptions, and Definitions.'' \textit{Annual Review of Statistics and Its Applications} 2021: 8: 141 --163.
\item
Morgan, S.L. and C. Winship(2015) \textit{Counterfactuals and Causal Inference}, second edition. Cambridge University Press.
\item
Mullainathan, S. 2018. ``Biased Algorithms Are Easier to fix Than Biased People.'' \textit{New York Times} December 6, 2019. https://www.nytimes. com/2019/12/06/business/algorithm-bias-fix.html.
Muller, C. (2021) ``Exclusion and Exploitation: The Incarceration of Black Americans fro Slavery to the Present.'' \emph{Science} 374(6565): 282 -- 286.
\item
Mukherjee, D., Yurochkin, M., Bannerhee, M., and Sun, Y. (2020) ``Two simple Ways to Learn Individual Fairness Metrics from Data,'' \textit{Proceeding of Machine Learning Research} 119: 7097 -- 7107. 
\item
Nabi, R., Malinsky, D., and I. Shpitser (2019) ``Learning Optimal Fair Policies.''\textit{Proceedings of Machine Learning Reserch} 97: 464 -- 4682.
\item
Nath, S.V. (2006) ``Crime Pattern Detection Using Data Mining.'' \textit{2006 IEEE/WIC/ACM International Conference on Web Intelligence and Intelligent Agent Technology Workshops}, 2006, pp. 41-- 44.
\item
Ostrom, C.W., Ostrom, B.J., and M. Kleinman (2003) \textit{Judge and Discrimination: Assessing the Theory and Practice of Criminal Sentencing}. NCJRS, U.S. Department of Justice, Washington, D.C.
\item
Peyr\'e, G., and M. Cuturi, M. (2019) \textit{Computational Optimal Transport With Applications to Data Science}. NOW Publishers.
\item
Pooladian, A.-A., and J. Niles-Weed (2021) ``Entropic Estimation of Transport Maps.'' arXiv:2109.12004v1 [math.ST]
\item
Regents of University of California v. Bakke (1978) 438 U.S. 265, 313. 
\item
Robert Wood Johnson Foundation (2017) ``Discrimination in America: Experiences of Views of African Americans. ''https://media.npr.org/assets/
img/2017/10/23/discriminationpoll- african-americans.pdf
\item
Rocque, M. (2011) ``Racial Disparities in the Criminal Justice System and Perceptions of Legitimacy: A Theoretical Linkage.'' \textit{Race and Justice} 1(3): 292 -- 315.	
\item
Romano, Y., Barber, R.F., Sabatti, C., and E.J. Candes (2019) ``With Malice Toward None: Assessing Uncertainty via Equalized Coverage.'' axXIiv: 1908.05428v1 [stat, ME]
\item
Rothenberg, P.S. (2008)\textit{White Privilege} Catherine Woods. 
\item
Rucker, J.M., and J.A. Rocheson (2021) ``Toward an Understanding of Structural Racism: Implications for Criminal Justice,'' \emph{Science} 374 (6565): 286 -- 290. 
\item
Skeem, J., and C. Lowenkamp (2020) ``Using Algorithms to Address Trade-Offs Inherent in Predicting Recidivism.'' \textit{Behavioral Science and Law} 38: 259--278.
\item
Shafer, G., and V. Vovk (2008) ``A Tutorial on Conformal Prediction.'' \textit{Journal of Machine Learning Research} 9: 371 -- 421.
\item
Si, N., Murthy, K., Blanchet, J., and V.A. Nguyen (2021) ``Testing Group Fairness via Optimal Transport Projections.'' Proceedings of the 38th International Conference on Machine Learning, PMLR 139.
\item
Sorenson, S.B., Sinko, L., and R.A. Berk (2021) ``The Endemic Amid the Pandemic: Seeking Help for Violence against Women in the Initial Phases of COVID-19.'' \textit{Journal of Interpersonal Violence} published online March, 2021.  
\item
Starr, S.B. (2014) ``Evidence-Based Sentencing and the Scientific Rationalization of Discrimination. \textit{Stanford Law Review} 66: 803 -- 872.
\item
Stewart, E.A., Warren, P.Y., Hughes, C., and Brunson, R.K. (2020) ``Race, Ethnicity, and Criminal Justice Contact: Reflections for Future Research,'' \textit{Race and Justice} 10 (2): 119 --149.
\item
Tibshirani, R.J., Barber, R.F., Cand\`{e}s, E.J. and A. Ramdas (2020) ``Conformation Prediction Under Covariate Shift.'' arXiv: 1904.06019v3 [stat.ME].
\item
Thompson, W.C., and E.L. Schumann (1987) ``Interpretation of Statistical Evidence in Criminal Trials: The Prosecutor's Fallacy and the Defense Attorney's Fallacy.'' \emph{Law and Human Behavior} 11: 167 -- 187.
\item
Tonry, M. (2014) ``Legal and Ethical Issues in The Prediction of Recidivism.'' \textit{Federal Sentencing Reporter} 26(3): 167 -- 176.
\item
Van Cleve, N.G. and L. Mayes(2015) ``Criminal Justice Through “Colorblind” Lenses: A Call to Examine the Mutual Constitution of Race and Criminal Justice.'' \textit{Law \& Social Inquiry} 40(2): 406 -- 432.
\item
Vovk, V., Gammerman, A., and G. Shafer (2005), \textit{Algorithmic Learning in
a Random World}, NewYork: Springer
\item
Vovk,V., Nouretdinov, I., and A. Gammerman (2009), ``On-Line Predictive Linear Regression.'' \textit{The Annals of Statistics} 37: 1566 -- 1590.
\item
Wacquant, L. (2002) ``From Slavery to Mass Incarceration: Rethinking the `race question' in the US.'' \textit{New Left Review} 13:41 -- 73.
\item
Wallis, J. (2017) \textit{America's Original Sin} Brazos Press.
\item
Wyner, A.J., Olson, M., Bleich, J, and D. Mease (2015) ``Explaining the Success of AdaBoost and Random Forests as Interpolating Classifiers.'' 
\textit{Journal of Machine Learning Research} 18(1): 1--33.
\item
Yates, J. (1997) ``Racial Incarceration Disparity Among States.'' \textit{Social Science Quarterly} 78(4) 1001 -- 1010.
\item
Zafar, M.B., Martinez, I.V., Rodriguez, M.,B.,  and K. Gummadi. (2017) ``Fairness Constraints: A Mechanism for Fair Classification.'' In Proceedings of the 20th International Conference on Artificial Intelligence and Statistics (AISTATS). Fort Lauderdale, FL, 2017.
\item
Zemel, R., Wu, Y., Swersky, K., Pitassi, T., and C. Dwork (2013) ``Learning Fair Representations.'' \textit{Proceedings of Machine Learning Research} 28 (3) 325 -- 333.

\end{description}
\bibliography{references}
\bibliographystyle{apalike}
\end{document}